\documentclass[journal,onecolumn,draft]{IEEEtran}

\pdfoutput=1

\ifCLASSINFOpdf
\else
\fi

\usepackage{graphicx}
\usepackage{mathtools}
\usepackage{amsmath}
\usepackage{amsfonts}
\usepackage{enumerate}
\usepackage{dsfont}
\usepackage{amssymb}
\usepackage{bbold}
\usepackage{mathabx} %NEEDED THIS FOR WIDEBAR(?)
\usepackage{marginnote}

\newtheorem{theorem}{Theorem}[section]
\newtheorem{lemma}[theorem]{Lemma}
\newtheorem{proposition}[theorem]{Proposition}

\newenvironment{definition}[1][Definition:]{\begin{trivlist}
\item[\hskip \labelsep {\it #1}]}{\end{trivlist}}
\newenvironment{example}[1][Example:]{\begin{trivlist}
\item[\hskip \labelsep {\it #1}]}{\end{trivlist}}

\newcommand{\qed}{\nobreak \ifvmode \relax \else
      \ifdim\lastskip<1.5em \hskip-\lastskip
      \hskip1.5em plus0em minus0.5em \fi \nobreak
      \vrule height0.75em width0.5em depth0.25em\fi}

\usepackage[latin1]{inputenc}
\usepackage{tikz}
\usepackage{verbatim}
\usetikzlibrary{patterns}

\newif\iffinal % introduce a switch for draft vs. final document
\pgfrealjobname{DC-journal} % <-- NOTE: this needs to be the real document's basename
\iffinal
  
\else
  
\fi

\newcommand{\smthker}{w}

\newcommand{\altPhi}{\phi}
\newcommand{\altPhiSI}{\xi_\phi}
\newcommand{\ind}{{\mathbb{1}}}
\newcommand{\indicator}[1]{\ind_{\{ #1 \}}}
\newcommand{\cut}[2]{\lfloor #1 \rceil_{\!#2}}

\newcommand{\reals}{\mathds{R}}
\newcommand{\veq}{\doteqdot}
\newcommand{\tinyt}[2]{#1_{\text{\tiny #2}}} 

\begin{document}
\title{Displacement Convexity in Spatially \\Coupled Scalar Recursions}

\author{\IEEEauthorblockN{Rafah El-Khatib, Nicolas Macris, Tom Richardson, Ruediger Urbanke}\\
\IEEEauthorblockA{EPFL Switzerland, and Qualcomm USA\\
Emails: \{rafah.el-khatib,nicolas.macris,ruediger.urbanke\}@epfl.ch, tomr@qti.qualcomm.com}
}

\maketitle

\begin{abstract}
We introduce a technique for the analysis of general spatially coupled systems that are governed by scalar 
recursions. Such systems can be expressed in 
variational form in terms of a potential functional. We show, under mild conditions, that the potential 
functional is \emph{displacement convex} and that the minimizers are given by the fixed points of the recursions.
Furthermore, we give the conditions on the system such that 
the minimizing fixed point is unique up to translation along the spatial direction. The condition matches those in \cite{KRU12} for the existence of spatial fixed points. 
\emph{Displacement convexity} applies to a wide range of spatially coupled recursions appearing in coding theory,
compressive sensing, random constraint satisfaction problems, as well as statistical mechanical models. 
We illustrate it with applications to 
Low-Density Parity-Check and generalized LDPC codes used for transmission on the binary erasure channel,
or general binary memoryless symmetric channels within the Gaussian reciprocal channel approximation, as well 
as compressive sensing. 

\end{abstract} \IEEEpeerreviewmaketitle

\section{Introduction}
Spatially coupled systems have been used recently in various
frameworks such as coding \cite{Zigangirov}, \cite{lentmaier2005terminated}, \cite{lentmaier2010iterative}, \cite{kudekar2013spatially} 
(for a review of applications in the context of communications see \cite{kudekar2013spatially} and references therein), 
compressive sensing \cite{donoho2012information}, \cite{krzakala2012probabilistic}, statistical physics
\cite{hassani2010itw}, \cite{hassani2012meanfield}, 
and random constraint satisfaction problems \cite{hassani2013threshold}, \cite{achlioptas2016bounds}. These systems exhibit excellent
performance, often optimal, under low complexity message passing
algorithms, due to the threshold saturation phenomenon \cite{kudekar2013spatially}, \cite{yedla2014simple}, \cite{kumar2014threshold}. 
For example, spatially coupled high-degree regular LDPC codes achieve the Shannon capacity
under belief propagation \cite{kudekar2013spatially}, \cite{kumar2014threshold}. Another line of research has used spatially coupled 
constructions to prove results about the original uncoupled underlying model. For example, this idea was used to obtain proofs 
of replica-symmetric formulas for the mutual information in coding \cite{giurgiu2015spatial}, in rank-one matrix factorization
\cite{barbier2016mutual}, and to improve provable algorithmic lower bounds on phase transition thresholds of random constraint satisfaction 
problems \cite{achlioptas2016bounds}. 

Given the success of spatial coupling in a wide variety of problems, it should hardly come as a surprise that there are fundamental mathematical
structures behind spatially coupling. This paper is concerned with a somewhat hidden convexity structure called 
\emph{displacement convexity}. Some of our preliminary work on this matter appeared in \cite{el2013displacement}, \cite{ElKhatib}, \cite{el2014analysis}.

The large system asymptotic performance of spatially coupled systems is assessed
by the solutions of coupled density evolution (DE) type update equations. In 
general, the fixed points of these equations 
can be viewed as the stationary point equations of a functional 
that is typically called the 
``potential functional'' and is an ``average form'' of the Bethe free energy \cite{yedidia2001bethe} of the underlying graphical model.\footnote{
In the context of statistical mechanics, the potential functional is the ``replica free energy functional'' \cite{mezard1987spin}.
The precise connection between the Bethe free energy and the potential functional in the case of coding can be 
found in \cite{kumar2014threshold}.}
It has already been recognized that this variational 
formulation is a powerful tool to analyze DE updates under 
suitable initial conditions \cite{KRU12}, \cite{hassani2010itw}, \cite{yedla2014simple}, \cite{kumar2014threshold}.
There are various possible formulations of this 
potential functional; in this paper,  
we will use the representation from \cite{KRU12} for scalar systems. 

In a previous contribution \cite{el2013displacement}, we showed that the
potential, in the form given in \cite{yedla2014simple}, associated to a spatially
coupled low-density parity-check (LDPC) code whose single system
is the $(\ell,r)$-regular Gallager ensemble, with transmission over
the binary erasure channel with parameter $\epsilon$, or the BEC($\epsilon$), has a convex structure called \emph{displacement
convexity}. This structure is well-known in the theory of optimal
transport \cite{Villani}. In fact, the potential we consider in
\cite{el2013displacement} is \emph{not} convex in the usual sense but it
\emph{is} in the sense of displacement convexity. This, in itself,
is an interesting property.  
Although 
the formalism in \cite{el2013displacement} can be extended to more general scalar recursions, for example, those pertaining to 
irregular LDPC codes, it does not appear to extend to a very wide class of general scalar recursions.
The main purpose of the present paper is to prove that a rather general class
of scalar systems also exhibits the property of displacement convexity,
and even strict displacement convexity under rather mild assumptions. 
Although the analysis of the present paper is similar in spirit to \cite{el2013displacement} it is also significantly different and more far reaching in its range of applications. 
We use the potential in the representation of \cite{KRU12}
which allows to obtain much more general proofs that hold under quite mild conditions. The results are applicable 
to recursions appearing not only in coding, but also in compressive sensing and random constraint satisfaction problems. 

The main propositions of this paper are: Proposition~\ref{propConv} that states that the potential functional has the displacement convexity property; Proposition~\ref{prop:CFPmin} that asserts that monotonic minimizers of the potential functional are fixed point solutions of the 
spatially coupled DE equations (in a generalized sense); Proposition~\ref{prop:unicity} that gives the condition for the unicity of the minimizers up to translations along the spatial axis. It is also of interest that the potential functional satisfies a rearrangement inequality, namely Proposition~\ref{prop:mainRearrange} that ensures that one can find minimizers among monotonic spatial fixed points. The conditions for our results to hold are rather mild and essentially match those in \cite{KRU12} for the existence of spatial fixed points. 

This manuscript is organized as follows. Section~\ref{sectionSetUp}
introduces spatially coupled recursions and the variational formulation.  In
Section~\ref{sectionRearr}, we prove rearrangement inequalities that
allow us to reduce the search for minima of the potential to
a space of monotonic functions, and, in Section~\ref{sectionExistence}, we discuss the existence question using the direct
method from functional analysis. The potential is
shown to be displacement convex in Section~\ref{sectionDC}.  
In Section~\ref{sectionCFP}, we generalize the notion of fixed point solutions to
the DE equations and show that such generalized solutions are minimizers of
the potential.  Unicity of the minimizer is addressed
in Section~\ref{sectionUnicity}. In Section~\ref{SectionApplications}, we illustrate 
displacement convexity with
applications to coding and compressive sensing.

\section{Set Up and Variational Formulation} \label{sectionSetUp}

In this section, we explain the set-up for general spatially coupled scalar recursions and give 
a variational formulation of these recursions. The fixed point equations of the scalar recursions
will be generically called ``density evolution'' (DE) equations. The case of regular $(\ell,r)$-LDPC code ensembles
with transmission over the BEC$(\epsilon)$ will serve as a concrete running example
for the setting.

\begin{figure}
\centering
    \beginpgfgraphicnamed{plotEXITcurves-external}%
    \begin{tikzpicture}[scale=5]
% filling up areas - left to right

% area 1
\draw [fill=gray,fill opacity=0.2] plot [smooth,samples=100,domain=0:0.19/0.76](\x, {(sin(18*0.76*\x r)+20*0.76*\x)/19.08}) -- plot [smooth,samples=100,domain=0.19/0.76:0] (\x, {(-sin(0.76*18*\x r)+17*0.76*\x)/19.08});

% area 2
\draw [fill=gray,fill opacity=0.8] plot [smooth,samples=100,domain=0.19/0.76:0.32/0.76](\x, {(sin(18*0.76*\x r)+20*0.76*\x)/19.08}) -- plot [smooth,samples=100,domain=0.32/0.76:0.19/0.76] (\x, {(-sin(0.76*18*\x r)+17*0.76*\x)/19.08});

% area 3
\draw [fill=gray,fill opacity=0.2] plot [smooth,samples=100,domain=0.32/0.76:0.5/0.76](\x, {(sin(18*0.76*\x r)+20*0.76*\x)/19.08}) -- plot [smooth,samples=100,domain=0.5/0.76:0.32/0.76]  (\x, {(-sin(0.76*18*\x r)+17*0.76*\x)/19.08});
\draw [fill=gray,fill opacity=0.2] plot [smooth,samples=100,domain=0.5/0.76:0.55/0.76](\x, {(sin(18*0.76*\x r)+20*0.76*\x)/19.08}) -- plot [smooth,samples=100,domain=0.55/0.76:0.5/0.76] (\x, {(-sin(0.76*18*\x r)+17*0.76*\x)/19.08+(31*(0.76*\x-0.5)-sin(30*(0.76*\x-0.5) r))/19.08});

% area 4
\draw [fill=gray,fill opacity=0.8] plot [smooth,samples=100,domain=0.55/0.76:0.7/0.76](\x, {(sin(18*0.76*\x r)+20*0.76*\x)/19.08}) -- plot [smooth,samples=100,domain=0.7/0.76:0.55/0.76] (\x, {(-sin(0.76*18*\x r)+17*0.76*\x)/19.08+(31*(0.76*\x-0.5)-sin(30*(0.76*\x-0.5) r))/19.08});
\draw [fill=gray,fill opacity=0.8] plot [smooth,samples=100,domain=0.7/0.76:1](\x, {(sin(18*0.76*\x r)+20*0.76*\x+39*(0.76*\x-0.7)+sin(10*(0.76*\x-0.7) r))/19.08}) -- plot [smooth,samples=100,domain=1:0.7/0.76] (\x, {(-sin(0.76*18*\x r)+17*0.76*\x)/19.08+(31*(0.76*\x-0.5)-sin(30*(0.76*\x-0.5) r))/19.08});

%ticks
\foreach \x in {0,0.1,...,1}
        \draw (\x,1pt) -- (\x,-1pt); %{%\pgfmathprintnumber[precision=1]{\x} \\
%     \pgfmathprintnumber[format/precision=1]{\x}};
\foreach \y in {0,0.1,...,1}
        \draw (1pt,\y) -- (-1pt,\y);% node[anchor=north] %{%\pgfmathprintnumber[precision=1]{\y} \\
%     \pgfmathprintnumber[format/precision=1]{\y}};

% axes increments
\node (x0) at (0,-0.07) {$0$};
%\node (x1) at (0.1,-0.07) {$0.1$};
\node (x2) at (0.2,-0.07) {$0.2$};
%\node (x3) at (0.3,-0.07) {$0.3$};
\node (x4) at (0.4,-0.07) {$0.4$};
%\node (x5) at (0.5,-0.07) {$0.5$};
\node (x6) at (0.6,-0.07) {$0.6$};
%\node (x7) at (0.7,-0.07) {$0.7$};
\node (x8) at (0.8,-0.07) {$0.8$};
%\node (x9) at (0.9,-0.07) {$0.9$};

\node (y0) at (-0.09,0) {$0$};
%\node (y1) at (-0.09,0.1) {$0.1$};
\node (y2) at (-0.09,0.2) {$0.2$};
%\node (y3) at (-0.09,0.3) {$0.3$};
\node (y4) at (-0.09,0.4) {$0.4$};
%\node (y5) at (-0.09,0.5) {$0.5$};
\node (y6) at (-0.09,0.6) {$0.6$};
%\node (y7) at (-0.09,0.7) {$0.7$};
\node (y8) at (-0.09,0.8) {$0.8$};
%\node (y9) at (-0.09,0.9) {$0.9$};

\node (x) at (1.05,0) {$\mathbf{u}$};

%contour
\draw[ultra thick, <->] (0,1) -- (0,0) -- (1,0);
\draw[] (1,0) -- (1,1) -- (0,1);

% lines to cover contour
\draw[gray!20!white, ultra thick] (0.5/0.76,0.428) -- (0.5/0.76,0.5447);
\draw[gray!80!white, ultra thick] (0.7/0.76,0.744) -- (0.7/0.76,0.9615);
\draw[gray!80!white, ultra thick] (0.927,0.91) -- (0.927,0.9615);

% grid
\draw[step=0.1,gray,thin] (0,0) grid (1,1);

% blue curve
\draw[blue,thick, domain=0:0.7/0.76] plot (\x, {(sin(18*0.76*\x r)+20*0.76*\x)/19.08});
\draw[blue,thick, domain=0.7/0.76:1] plot (\x, {(sin(18*0.76*\x r)+20*0.76*\x+39*(0.76*\x-0.7)+sin(10*(0.76*\x-0.7) r))/19.08});

% red curve
\draw[red,thick, domain=0:0.5/0.76] plot (\x, {(-sin(0.76*18*\x r)+17*0.76*\x)/19.08});
\draw[red,thick, domain=0.5/0.76:1] plot (\x, {(-sin(0.76*18*\x r)+17*0.76*\x)/19.08+(31*(0.76*\x-0.5)-sin(30*(0.76*\x-0.5) r))/19.08});

% function labels
\node (hf) [red, scale=1] at (0.7,0.85) {$h_f(u)$};
\node (hgInv) [blue,scale=1] at (0.85,0.5) {$h_g^{-1}(u)$};

% axes
\draw (0,0) -- coordinate (x axis mid) (1,0);
\draw (0,0) -- coordinate (y axis mid) (0,1);

\end{tikzpicture}%
    \endpgfgraphicnamed%
  
\caption{\label{plot} A generic example of the systems we consider. The EXIT-like curves are $h_f$ (in red) and $h_g^{-1}$ (in blue). The signed area $A(h_f,h_g;1)$ from~\eqref{eqnSignedArea} is the sum of the light gray areas (positively signed) and the dark gray areas (negatively signed), and it is equal to $0$.}
\end{figure}
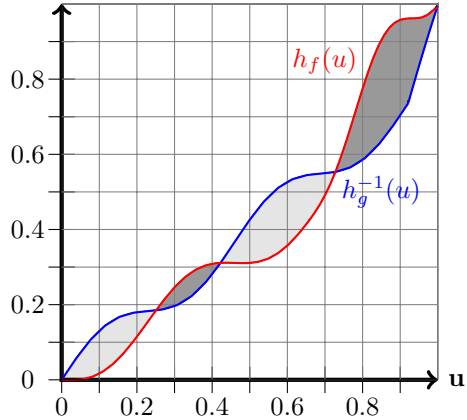

Consider the pair of DE fixed point equations 
\begin{align}\label{simple-DE}
 \begin{cases}
 u=h_g(v), \\
 v=h_f(u),
  \end{cases}
\end{align}
where $u,v\in [0,1]$. The {\it update functions} $h_f$, $h_g$ are 
assumed to be non-decreasing from $[0,1]$
to $[0,1]$, and normalized such that $h_f(0)=h_g(0) =0$ and $h_f(1)=h_g(1)=1$.
We will think of them as EXIT-like curves of DE $(u,h_f(u))$ and $(h_g(v),v)$
for $u,v\in [0,1]$ (see Fig.~\ref{plot}). It is always
possible to adopt this normalization in specific applications.

\begin{example}
Take an $(\ell,r)$-regular Gallager ensemble, 
with transmission over the BEC($\epsilon$).
Let $\mathtt{y}$ (resp. $\mathtt{x}$) be the erasure probability emitted by the check (resp.  variable) nodes.
The DE fixed point equations are $\mathtt{y} = 1-(1-\mathtt{x})^{r-1}$ and  $\mathtt{x} = \epsilon \mathtt{y}^{\ell -1}$. In this paper, we are interested in the specific 
value $\epsilon=\epsilon_{\text{\tiny MAP}}$
which is the MAP threshold of the ensemble. Let $\mathtt{x}_{\text{\tiny MAP}}$, $\mathtt{y}_{\text{\tiny MAP}}$ be the non-trivial {\it stable} fixed point 
when $\epsilon=\epsilon_{\text{\tiny MAP}}$. 
To achieve the normalization of~\eqref{simple-DE} we make the change of variables $\mathtt{y} = \mathtt{y}_{\text{\tiny MAP}} u$ and $\mathtt{x}=\mathtt{x}_{\text{\tiny MAP}} v$, so that the DE equations become $u = \mathtt{y}_{\text{\tiny MAP}}^{-1}(1 - (1-  \mathtt{x}_{\text{\tiny MAP}} v)^{r-1})$ and $v=\epsilon_{\text{\tiny MAP}} \mathtt{x}_{\text{\tiny MAP}}^{-1}\mathtt{y}_{\text{\tiny MAP}}^{\ell -1} u^{\ell -1}$.  Note that we must have $1=\mathtt{y}_{\text{\tiny MAP}}^{-1}(1 - (1-  \mathtt{x}_{\text{\tiny MAP}} )^{r-1})$ and $1=\epsilon_{\text{\tiny MAP}} \mathtt{x}_{\text{\tiny MAP}}^{-1}\mathtt{y}_{\text{\tiny MAP}}^{\ell -1}$.
We then set 
\begin{align}
\begin{cases}
h_g(v) = \mathtt{y}_{\text{\tiny MAP}}^{-1}(1 - (1 - \mathtt{x}_{\text{\tiny MAP}} v)^{r-1}),\\
h_f(u) = u^{\ell -1},
\end{cases}
\end{align}
which satisfy the required normalizations $h_f(0)=h_g(0)=0$ and $h_f(1)=h_g(1)=1$. The corresponding EXIT curves 
have three intersections. The one at $(0,0)$ corresponds to the trivial fixed point  of DE, the one at $(1,1)$ corresponds to the 
stable non-trivial fixed point of DE, and the third one at a middle point  corresponds to the unstable fixed point.  
\end{example}

The natural setting for displacement convexity, at least in the context of spatial coupling, is the continuum
setting, which can be thought of as an approximation of the corresponding discrete
system in the regime of large spatial length and coupling window size. The
continuum limit has already been introduced in the literature as a
convenient means to analyze the behavior of an originally discrete
model \cite{KRU12}, \cite{donoho2012information}, \cite{hassani2010itw}.

Consider a spatially coupled system with an averaging window
$w:\mathds{R}\to \mathds{R}$ which is always assumed to be
bounded, non-negative, even, integrable, and normalized such that
$\int_{\mathds{R}} {\mathrm d}x\, w(x) =1.$ 
The averaging window is the means for the ``coupling'' in ``spatial coupling''.
Let us define the constant
\begin{align}\label{eqn:Cwbound}
C_\smthker := \int_\reals {\mathrm d}x  \, |x| \smthker (x).
\end{align}
We assume throughout the paper that $C_\smthker$ is finite. 
As we shall see, this is directly related to finiteness of the potential. 
Let $f, g:\mathds{R}\to [0,1]$ be two functions 
and denote by $f^w=f \otimes w$ and $g^w=g \otimes w$ their usual convolutions 
with $w$, i.e., $f^w(x) = \int_{\mathds{R}} {\mathrm d}x\, f(y) w(x-y)$ and $g^w(x) = \int_{\mathds{R}} {\mathrm d}x\, g(y) w(x-y)$.
The pair of fixed point DE equations of a spatially coupled scalar continuous system are
\begin{align}\label{eqn:DE}
\begin{cases}
g(x)=h_g(f^w(x)),\\
f(x)=h_f(g^w(x)),
\end{cases}
\end{align}
where $x\in \mathds{R}$ is the spatial position.  We will often refer to the functions $f$, $g$ as {\it
profiles} and to $h_f$, $h_g$ as {\it update functions}.
A pair of profiles $f,g: \mathds{R}\to [0,1]$ that solves the above equations almost everywhere will
be called a {\it fixed point}, FP for short. Note that~\eqref{eqn:DE} are \emph{non-local} equations because of the coupling through $w.$

In this paper, we are interested in profiles  $p:\mathds{R}\rightarrow[0, 1]$  ($p$ denotes a generic profile like $f$ and $g$) that satisfy
the limit conditions
\begin{equation}\label{eqn:RearrLimits}
\lim\limits_{x\rightarrow -\infty}p(x)=0,\qquad \lim\limits_{x\rightarrow +\infty}p(x)=1.
\end{equation}
We note that these two limit values are the extreme fixed points of~\eqref{simple-DE}
We will refer to such profiles as {\em interpolating profiles}.
A pair $f,g$ of interpolating profiles that solves~\eqref{eqn:DE} is called an
{\em interpolating FP}. 

\begin{definition}
A function $p:\mathds{R}\rightarrow[0, 1]$ satisfying~\eqref{eqn:RearrLimits} is called an interpolating profile.
A pair $f,g$ of interpolating profiles that solves~\eqref{eqn:DE} almost everywhere, i.e., up to a set of measure zero,  is called an interpolating fixed point (FP).
\end{definition}

In Section~\ref{sectionRearr}, we show that when minimizing the potential 
functional over the space of interpolating profiles we can focus 
on monotonic (non-decreasing) profiles. 

\subsection{Potential function associated to~\eqref{simple-DE}}\label{subsec:potfunctions}

In \cite{KRU12} the following potential function is introduced,
\begin{align}\label{simple-pot}
\phi(h_f,h_g;u,v) =  \int_0^u \text{d}u'\,h_g^{-1}(u') +  \int_0^v \text{d}v' \, h_f^{-1}(v') \, - uv\,.
\end{align}
Often, when they are clear from context or irrelevant, 
we will drop the update functions $h_f$ and $h_g$ as arguments from the notation and denote this potential function by $\phi(u,v)$. 
Since $h_g^{-1}$ and $h_f^{-1}$ are non-decreasing the potential $\phi(u,v)$ is convex in $u$ for fixed $v$ and convex in $v$ for fixed $u.$
It is minimized over $v$ by setting $v=h_f(u)$ 
and over $u$ by setting $u=h_g(v)$.

Substituting  $v=h_f(u)$ in~\eqref{simple-pot}, we obtain the
integral of the \emph{signed area} between the two EXIT curves (see fig.~\ref{plot}) as
\begin{align}
 A(h_f, h_g; u) &= \phi(h_f,h_g;u,h_f(u))\nonumber\\
 & = \int_0^u \mathrm{d}u^\prime \, (h_g^{-1}(u^\prime) - h_f(u^\prime)).\label{eqnSignedArea}
\end{align}
Note that this is the signed area bounded by the two curves and the region 
between the vertical axis at the origin and a vertical axis at $u$.

In \cite{KRU12}, the following key result was shown.  It states that for an interpolating FP to exist the potential $\phi$ must be minimal at both limit points.
\begin{lemma}\label{lem:PGC}
If there exists an interpolating FP solution to~\eqref{eqn:DE}, then
$\phi(h_f,h_g;u,v) \ge 0$ for all $u,v\in [0,1]$  and
$A(h_f, h_g;1) = \phi(h_f,h_g;1,1) = 0.$
\end{lemma}

The result applies not only to interpolating FPs but also to a relaxed definition
of interpolating ``consistent" FPs (CFPs) that we define in Section~\ref{sectionCFP}.
In \cite{KRU12}, when the assumption $\phi(h_f,h_g;1,1) = 0$ is made, the condition 
$\phi(h_f,h_g;u,v) \ge 0$ for all $u,v\in [0,1]$  
is termed the \emph{positive gap condition} (PGC). 
In this paper we will additionally assume $\phi(h_f,h_g;1,1) = 0$ throughout so the term positive gap condition will be 
used to imply both this equality and the inequality in Lemma~\ref{lem:PGC}.

When the inequality in Lemma~\ref{lem:PGC} is strict, i.e., $\phi(h_f,h_g;u,v) > 0$ for $(u,v) \not\in \{(0,0),(1,1)\}$
then the condition is termed the \emph{strictly positive gap condition} (SPGC) in \cite{KRU12}.
In this case, it was shown that an interpolating fixed point profile exists
provided $\smthker$ is strictly positive
on the interior of some interval $[-W,W]$ and zero off of the interval.
This support condition on $\smthker$ can be relaxed under various other
conditions (see \cite{KRU12}).

\begin{definition}
We say that the positive gap condition (PGC) is satisfied when $\phi(h_f,h_g;1,1) = 0$ and $\phi(h_f,h_g;u,v) \ge 0$ for all $u,v\in [0,1]$.
The strictly positive gap condition is satisfied when $\phi(h_f,h_g;1,1) = 0$ and $\phi(h_f,h_g;u,v) > 0$ for $(u,v) \not\in \{(0,0),(1,1)\}$.
\end{definition}

\begin{example}
 For the $(\ell,r)$-regular Gallager ensemble, 
with transmission over the BEC($\epsilon$) with $\epsilon=\epsilon_{\rm MAP}$ we have the potential function
\begin{align*}
\phi(u,v) = &
 \frac{1}{\tinyt{\mathtt{x}}{MAP}}\Big\{u
-  \frac{r-1}{\tinyt{\mathtt{y}}{MAP} r} (1 - (1- \tinyt{\mathtt{y}}{MAP}u )^{\frac{r}{r-1}}))\Big\}
+\frac{\ell -1}{\ell}v^{\frac{\ell}{\ell -1}}-uv,
\end{align*}
and the signed area
\begin{align*}
A(h_f, h_g; u) = & \frac{1}{\tinyt{\mathtt{x}}{MAP}}
 \Big\{u
+  \frac{r-1}{\tinyt{\mathtt{y}}{MAP}r} ((1- \tinyt{\mathtt{y}}{MAP} u )^{\frac{r}{r-1}} -1)\Big\}
-\frac{u^{\ell}}{\ell}.
\end{align*}
Moreover, we have $A(h_f, h_g; 1) = 0$. In fact, this last constraint together with the two fixed point equations 
$\mathtt{y}_{\text{\tiny MAP}} = 1 - (1-  \mathtt{x}_{\text{\tiny MAP}} )^{r-1}$ and 
$\mathtt{x}_{\text{\tiny MAP}} = \epsilon_{\text{\tiny MAP}} \mathtt{y}_{\text{\tiny MAP}}^{\ell -1}$ completely determine 
$\epsilon_{\text{\tiny MAP}}$, $\mathtt{x}_{\text{\tiny MAP}}$ and $\mathtt{y}_{\text{\tiny MAP}}$. The SPGC holds for this example
(see Section~\ref{SectionApplications} for further illustration).
\end{example}

\subsection{Potential functional of the spatially coupled system~\eqref{eqn:DE}}

The solutions of spatially coupled DE equations~\eqref{eqn:DE} are given by the 
stationary point of a {\it potential functional} $\mathcal{W}$ of $f$ and $g$ defined below. 
This can be checked by setting the functional derivatives of this potential functional with respect to each of $f$ and $g$ to zero. We set
\begin{align}\label{eqnPotBothFts}
\mathcal{W}(f,g)=\int_{\mathds{R}}\mathrm{d}x\,I_{f,g,\smthker}(x),
\end{align}
where we have introduced the notation
\begin{align}
I_{f,g,\smthker}(x)= & \int_0^{g(x)}\mathrm{d}u\, h_g^{-1}(u)
+\int_0^{f(x)}\mathrm{d}v\, h_f^{-1}(v) 
-f^w(x)g(x).\label{eqn:Idef}
\end{align}
\begin{example}
For the $(\ell,r)$-regular LDPC code and transmission over the BEC($\epsilon$), the potential~\eqref{eqnPotBothFts} is
\begin{align*}
\mathcal{W}&(f,g) 
= \int\limits_{\mathds{R}}\mathrm{d}x\,\Big[\frac{1}{\mathtt{x}_{\text{\tiny MAP}}}\big\{g(x)
- \frac{r-1}{\tinyt{\mathtt{y}}{MAP} r} (1 - (1- \tinyt{\mathtt{y}}{MAP} g(x) )^{\frac{r}{r-1}})\big\}
+ \frac{\ell -1}{\ell} f(x)^{\frac{\ell}{\ell -1}}-f^w(x)g(x)\Big].
\end{align*}
\end{example}

Note that the limit of the integrand in~\eqref{eqnPotBothFts} (and the example) vanishes when $x\to -\infty$ because of the condition~\eqref{eqn:RearrLimits} on the profiles. It also vanishes when $x\to +\infty$ because of~\eqref{eqn:RearrLimits} and $A(h_f, h_g; 1)=0$. However, this 
does not suffice for the existence of the integral,
essentially due to the
fact that $f^\smthker - f$ may not be Lebesgue integrable (for monotonic profiles
this difficulty does not arise). So it is possible that ${\cal W}(f,g)$ fails to be well-defined as a Lebesgue integral for some choices of the interpolating profiles.

Once we consider interpolating profiles and assume the PGC and that $C_\smthker<\infty$,
we can circumvent this technical issue by defining the potential functional as 
\begin{align}\label{eqn:limitWdefinition}
{\cal W}(f,g) = \lim_{A,B\rightarrow\infty}\int_{-A}^{B} {\mathrm d}x I_{f,g,\smthker}(x).
\end{align}
We show below that the limit always exists (it is possibly $+\infty$). 

\begin{lemma}\label{lem:posfatou}
Assuming the PGC, we have for any interpolating profile pair $f,g$ that
\begin{align}\label{eqn:Wpositive}
{\cal W}(f,g) \ge \int_\reals {\mathrm d}x\, \phi(f^\smthker (x),g(x)),
\end{align}
and given a sequence of interpolating pairs
$f_i,g_i$ converging pointwise almost everywhere to an interpolating
pair $f,g$ we have
\begin{align}\label{eqn:Wfatou}
\liminf_{i\rightarrow \infty} {\cal W}(f_i,g_i) \ge {\cal W}(f,g).
\end{align}
\end{lemma}

\begin{IEEEproof}
Define $H_f(f)= \int_0^f {\mathrm d}v \,h_f^{-1}(v)$
and  $H_g(g)= \int_0^g {\mathrm d}u \,h_g^{-1}(u)$.   
Note that 
\begin{equation*}
I_{f,g,\smthker}=(H_g \circ g) +(H_f \circ f) -f^\smthker g.
\end{equation*}
Now, if we define 
\begin{equation*}
\tilde{I}_{f,g,\smthker}=(H_g \circ g) +(H_f \circ f)^\smthker -f^\smthker g ,
\end{equation*}
then 
\begin{align*}
\int_{-A}^B {\mathrm d}x\, I_{f,g,\smthker}(x) 
& =  \int_{-A}^B {\mathrm d}x\, \tilde{I}_{f,g,\smthker}(x) 
+ \int_{-A}^B {\mathrm d}x\, (I_{f,g,\smthker}(x)  - \tilde{I}_{f,g,\smthker}(x))
\nonumber \\ &
= \int_{-A}^B {\mathrm d}x\, \tilde{I}_{f,g,\smthker}(x) 
+ \int_{-A}^B {\mathrm d}x ((H_f \circ f) - (H_f \circ f)^\smthker(x)).
\end{align*}
Taking limits $A, B\to +\infty$ by definition~\eqref{eqn:limitWdefinition} and Lemma~\ref{lem:limintegral}, we obtain
\begin{align*}
{\cal W}(f,g) = \lim_{A, B\to +\infty}\int_{-A}^B {\mathrm d}x\, \tilde{I}_{f,g,\smthker}(x) .
\end{align*}
We will shortly see that the PGC implies $\tilde{I}_{f,g,\smthker}(x)$ is non-negative so that  
${\cal W}(f,g)$ is well defined (it is possibly $+\infty$). This also means that it is possible to adopt
\begin{equation}\label{eqn:altWdefinition}
{\cal W}(f,g) = \int_\reals {\mathrm d}x\, \tilde{I}_{f,g,\smthker}(x) ,
\end{equation}
as an alternative expression for ${\cal W}(f,g)$.

Now, note that $H_f(f)$ and $H_g(g)$ 
are convex functions because $h_f^{-1}$ and $h_g^{-1}$ are non-decreasing. Indeed
\begin{align*}
H_f(f+a) -H_f(f) & = \int_f^{f+a} {\mathrm d}v \, h_f^{-1}(v) 
\geq   a h_f^{-1}(f) 
=  a H_f^\prime(f).
\end{align*}
By Jensen's inequality we have
$$
(H_f \circ f)^\smthker \ge (H_f \circ f^\smthker),
$$
and we therefore obtain
\begin{align}\label{posi}
\tilde{I}_{f,g,\smthker}(x)\ge \phi(f^\smthker(x),g(x)),
\end{align}
which proves the non-negativity of $\tilde{I}_{f,g,\smthker}(x)$ since $\phi(f^\smthker(x),g(x))$ is non-negative 
by the PGC. 

Integrating~\eqref{posi} and using~\eqref{eqn:altWdefinition}, we obtain the first claim
\eqref{eqn:Wpositive} of the lemma. Furthermore, we get the second claim~\eqref{eqn:Wfatou}
directly by applying Fatou's lemma to~\eqref{eqn:altWdefinition} (we can apply Fatou's lemma since by~\eqref{posi}
$\tilde{I}_{f_i,g_i,\smthker}$ is a non-negative sequence, and it converges to $\tilde{I}_{f,g,\smthker}$).
\end{IEEEproof}

Let us remark that in the process of proving this lemma, we have seen  $\mathcal{W}(f,g)$ can be defined as~\eqref{eqn:limitWdefinition} or equivalently as~\eqref{eqn:altWdefinition}, as long as we assume the PGC, interpolating profiles and $C_w<+\infty$. 

\subsection{Discussion}

In Section~\ref{sectionCFP}, we show that among all interpolating profiles, monotonic 
interpolating CFPs yield minimizers of $\mathcal{W}$. To do that, we use 
rearrangement properties that are summarized in Section~\ref{sectionRearr}. 
%It is easy to see that 
For a fixed $f$, we always have
$\mathcal{W}(f,g) \ge \mathcal{W}(f,h_g \circ f^\smthker)$. This is because $I_{f,g,\omega}(x)$ is convex in $g(x)$ for fixed $f(x)$ and 
setting 
$g(x) = h_g(f^\smthker(x))$ minimizes $I_{f,g,\omega}(x)$ over $g(x)$ for fixed $f(x)$.

One of the main results of this paper is to show the {\em displacement convexity} of $\mathcal{W}$ in its two arguments.
More precisely, we can think of interpolating between two pairs $(f_0,g_0)$ and $(f_1,g_1)$ of
{\em monotonic} profiles by interpolating their inverse functions.
Hence, we consider
\begin{align*}
\begin{cases}
f_\lambda^{-1} &=  (1-\lambda) f_0^{-1}+  \lambda f_1^{-1}, \\
g_\lambda^{-1} &=  (1-\lambda) g_0^{-1}+  \lambda g_1^{-1} ,
\end{cases}
\end{align*}
and show that $\mathcal{W}(f_\lambda,g_\lambda)$ is a convex function of $\lambda.$ 
Note that for a monotonic interpolating profile $p$ the inverse function $p^{-1}(u)$ is uniquely defined for
almost all $u\in (0,1)$ and right and left limits $p^{-1}(u+)$ and $p^{-1}(u-)$, respectively, are uniquely determined.
Displacement convexity is explained 
in more detail in Section~\ref{sectionDC}. 
%Since we are interested in minimizers of $\mathcal{W}$ this property is significant.

Displacement convexity applies only to monotonic profiles.  In the next section, we address
the conditions under which one can conclude that minimizers of $\mathcal{W}$ 
satisfying~\eqref{eqn:RearrLimits} can taken to be monotonic.

The following quantities will play a crucial role in the remainder of this work,
\begin{align}
\Omega(x)=\int_{-\infty}^x\mathrm{d}z\,w(z), \qquad
V(x)=\int_{-\infty}^x\mathrm{d}z\,\Omega(z).\label{eqnDefV}
\end{align}
Here, $V$ is called {\it the kernel} for reasons that will become clear.
As will be seen, displacement convexity arises from the convexity of $V$.

\begin{lemma}\label{lemVconvex}
Assume that $C_\smthker<\infty$. Then, $V$ is well defined and convex.
\end{lemma}

\begin{IEEEproof}
Using integration by parts, we can write
\begin{align}
V(x)=\int_{-\infty}^{x}\mathrm{d}z \, \Omega(z)=x\,\Omega(x)\Big\vert_{-\infty}^x-\int_{-\infty}^{x} \mathrm{d}z \, zw(z).
\label{eqn:Vdecompose}
\end{align}
For $z\leq0$, we have
\begin{align*}
\int_{-\infty}^z \mathrm{d}x \, |x|w(x) \geq \int_{-\infty}^z \mathrm{d}x \, |z|w(x) = |z|\Omega(z) \geq 0,
\end{align*}
so taking $z\to -\infty$ shows $\lim_{z\to -\infty}z\,\Omega(z)=0$. Using~\eqref{eqn:Vdecompose} we conclude  that 
\begin{align}
 V(x) = x\,\Omega(x) - \int_{-\infty}^{x} \mathrm{d}z \, zw(z).
\end{align}
Thus, $V$ is finite and well-defined. Convexity follows because $V^{\prime\prime}(x) = w(x) \geq 0$.
\end{IEEEproof}

Much of the analysis in this paper proceeds relatively simply under
the assumption that 
$$
\int_\reals \mathrm{d}x\, (1-f^\smthker(x))g(x) < \infty.
$$
Most of our results will first be established under this assumption.
In general, however, this assumption is not needed and it is sufficient only that $C_\smthker < \infty.$
We typically generalize our results to this case by taking limits. Let us discuss this issue.

\begin{figure}
\centering
\includegraphics[draft=false,scale=1]{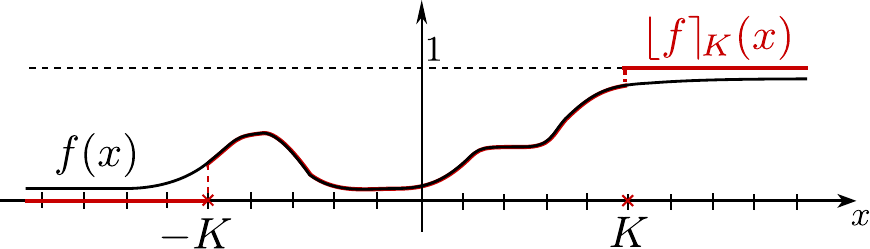}
\caption{A profile $f$ and its saturated version $\cut{f}{K}$.}
\label{saturation-example}
\end{figure}

\begin{definition}
We say that a function $f$ is {\em saturated} off of the
finite interval $[-K,K]$  
if $f(x)=0$ for  $x\in (-\infty,-K)$ and
$f(x)=1$ for $x \in (K,\infty).$
\end{definition}
Given a profile $f$ let us define $\cut{f}{K}$ by
\[
\cut{f}{K}(x) =\indicator{|x|\le K} f(x) + \indicator{x > K}\,.
\]
By definition, $\cut{f}{K}$  is saturated off of $[-K,K]$ (see Fig.~\ref{saturation-example}).

\begin{lemma}\label{lem:klimW} 
Let  $f,g$ be  interpolating profiles
and assume the PGC  and that $C_\smthker < \infty,$ then
\[
\lim_{K\rightarrow \infty}
\mathcal{W}(\cut{f}{K},\cut{g}{K}) =
\mathcal{W}(f,g)\,.
\] 
\end{lemma} 
\begin{IEEEproof}
See Appendix~\ref{app:basicbounds}.
\end{IEEEproof}

We end this section with another useful definition.
\begin{definition}
Assuming it exists, we define 
\begin{align}
& L(f,g) = \mathcal{W}(f ,g ) - \int_{\reals} (1-f^{\smthker}  (x)) g (x)\,{\mathrm d}x\,
=
\int_{\mathds{R}}&\mathrm{d}x
\Biggl(
\int_0^{f  (x)}\mathrm{d}v\, h_f^{-1}(v) 
-\int_0^{g  (x)}\mathrm{d}u\, (1-h_g^{-1}(u))
\Biggr).\label{eqn:Lfg}
\end{align}
\end{definition}
As we will see, the functional $L(f,g)$ captures the ``simple'' (uncoupled) part of $\mathcal{W}:$
It is invariant under increasing rearrangements and linear under displacement interpolation.

\begin{comment}
Note that $\mathcal{W}$ depends on $\smthker.$  Note that if, formally, $\smthker$ is $\delta,$ the
Dirac delta function then
\(
I_{f,g,\delta}(x) = \altPhi(g(x),f(x))
\) and so
\[
\mathcal{W}_\delta (f,g) = \int_{\reals}\mathrm{d}x\,\altPhi(g(x),f(x))\, \,.
\]
By the positive gap condition
the integrand is non-negative so
\(
\mathcal{W}_\delta(f,g) \ge 0\,.
\)
\end{comment}

\section{Rearrangements}\label{sectionRearr}

Displacement convexity is usually defined on  a space 
 of probability measures. For measures on the real line, 
 it is most convenient to view displacement 
 convexity on a space of cumulative distribution functions (cdf's). It is therefore fortunate that 
 the search for the global minimum of the 
 potential functional~\eqref{eqnPotBothFts} can be reduced to the space of 
 profiles $f$ and $g$ that are \emph{non-decreasing}.
In this section, we use the tool of \emph{increasing rearrangements} to show that such rearrangements of $f$ and $g$ can only
decrease the potential. 

\begin{figure}
\centering
\includegraphics[draft=false,scale=1]{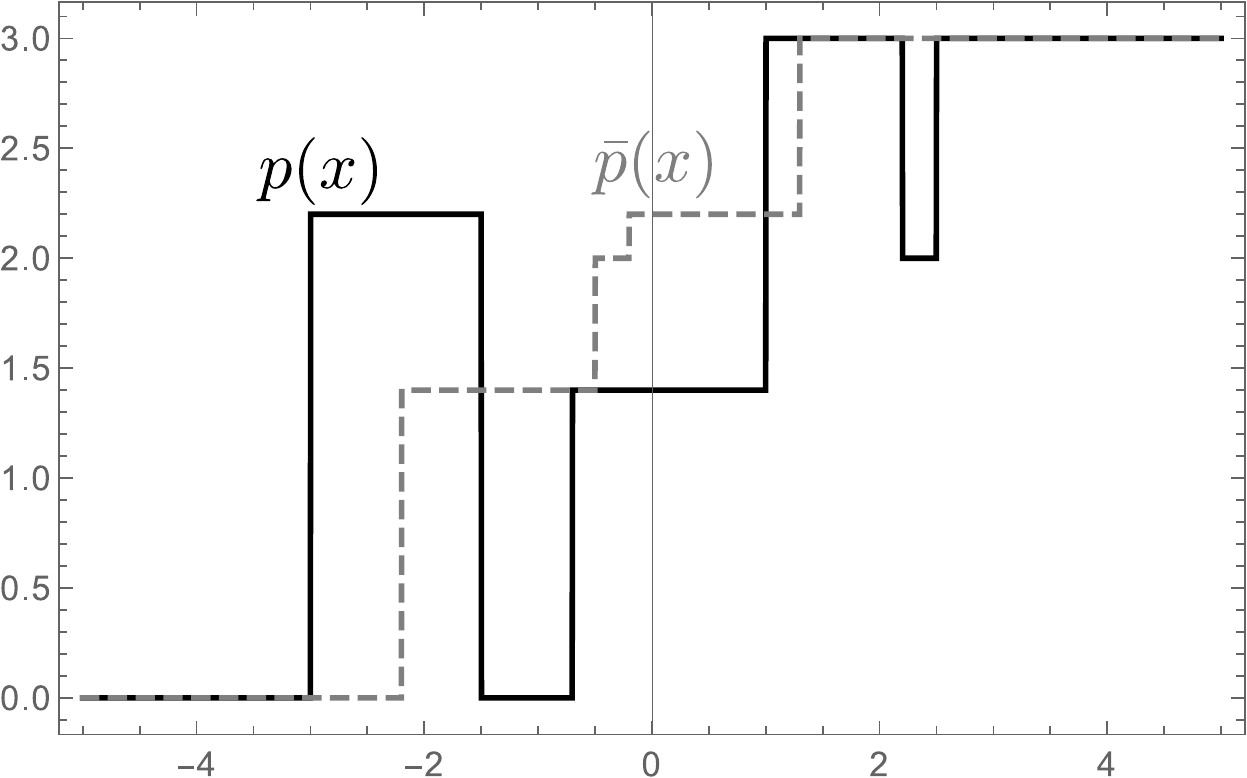}
\caption{A simple example of an increasing rearrangement for step functions.}
\label{rearrangement-example}
\end{figure}

Symmetric decreasing rearrangements are a classical tool in analysis, see \cite{Hardy}. Here we will use a closely related cousin namely increasing rearrangements (see \cite{Alberti}). Our presentation is self-contained and no previous exposure to rearrangements is needed.
Consider a profile $p:\mathds{R}\rightarrow[0, 1]$ 
that satisfies
\eqref{eqn:RearrLimits}.
The {\it increasing rearrangement\footnote{Note that an increasing rearrangement is not necessarily strictly increasing.}} of $p$ is the increasing function
$\bar{p}$ that has the same limits, and where the mass of each level set is in some sense preserved (here the mass of a level set
is infinite). More formally, let us represent  $p$ 
 in layer cake form as
\begin{align}\label{eqnLayerCake}
 p(x) =\int_{0}^{p(x)}\mathrm{d}t = \int_{0}^{1}\mathrm{d}t\, \ind_{E_t}(x),
\end{align}
where $\ind_{E_t}$ is the indicator function of the level
set $E_t = \{x\,\vert\, p(x) > t\}$.  For each value $t\in[0,1)$, the level set $E_t$
can be written as the disjoint union of a bounded set $A_t$ and a half line
$(a_t, +\infty)$. We define the rearranged set $\bar E_t = ( a_t- \vert
A_t\vert, +\infty)\,,$ and then
\begin{equation}
\bar p(x) = \int_{0}^{1} \mathrm{d}t\,  {\ind_{\bar {E_t}}}(x).
\end{equation}
A simple example capturing the notion of increasing rearrangement is shown in Fig.~\ref{rearrangement-example}.

\begin{lemma}\label{lem:pqinvar}
Let $p$ and $q$ be two profiles satisfying 
\eqref{eqn:RearrLimits}.
Then, assuming the left integral exists, we have
\[
\int_\reals \mathrm{d}x\, (p(x)-q(x)) 
=
\int_\reals \mathrm{d}x\, (\bar p(x)-\bar q(x))\,.
\]
\end{lemma}
\begin{IEEEproof}
For each $t\in (0,1)$ there exists a minimal $a_t$ such that
$(a_t,\infty) \subset \{x:p(x)>t\} \cap \{x:q(x)>t\}.$
Define $B_{p,t} = \{x:p(x)>t\} \backslash\, (a_t,\infty)$
and $B_{q,t} = \{x:q(x)>t\} \backslash\, (a_t,\infty).$ We also define the 
same quantities for the rearranged profiles $\bar p$ and $\bar q$, namely $\bar a_t$, $B_{\bar p,t}$ and $B_{\bar q,t}$.
We show below that 
\begin{align}\label{BbarB}
|B_{p,t}|-|B_{q,t}|  =  |B_{\bar p,t}|-|B_{\bar q,t}|.
\end{align}
Equation~\eqref{BbarB}
gives the result since, using the layer cake representation, it  
follows that 
\begin{align*}
& \int_\reals \mathrm{d}x\, (p(x)-q(x)) 
=
\int_0^1 \mathrm{d}t (|B_{p,t}|-|B_{q,t}|)
= \int_0^1 \mathrm{d}t (|B_{\bar p,t}|-|B_{\bar q,t}|)
= \int_\reals \mathrm{d}x\, (\bar p(x) - \bar q(x)) .
\end{align*}

Let us give an explicit argument for~\eqref{BbarB}. We note that the infinite part of a level set can only increase under an increasing
rearrangement, thus $(a_t, \infty)\subset (\bar a_t, \infty)$. So $(a_t, \infty)$ is common to 
$\{x:\bar p(x)>t\}$ and $\{x:\bar q(x)>t\}$ and subtracting it leaves two finite sets with the same finite measure since rearrangements
are measure preserving, i.e.,  
$$
\vert \{x:\bar p(x)>t\} \backslash\, (a_t,\infty)\vert = \vert\{x:p(x)>t\} \backslash\, (a_t,\infty)\vert,
$$
(with the {\it same} $a_t$ on both sides). Thus,
\begin{align*}
 \vert B_{\bar p, t}\vert & = \vert \{x:\bar p(x)>t\} \backslash\, (\bar a_t,\infty)\vert
 \nonumber \\ &
 = \vert \{x:\bar p(x)>t\} \backslash\, (\bar a_t,\infty)\vert - \vert(\bar a_t, a_t)\vert
 \nonumber \\ &
 = \vert \{x:\bar p(x)>t\} \backslash\, (a_t,\infty)\vert - \vert(\bar a_t, a_t)\vert
 \nonumber \\ &
 = \vert B_{p,t}\vert - \vert(\bar a_t, a_t)\vert.
\end{align*}
Similarly, $\vert B_{\bar q, t}\vert = \vert B_{q, t}\vert - \vert(\bar a_t, a_t)\vert$, and~\eqref{BbarB} follows from 
these two identities.
\end{IEEEproof}

\begin{figure}
\centering
\includegraphics[draft=false,scale=0.8]{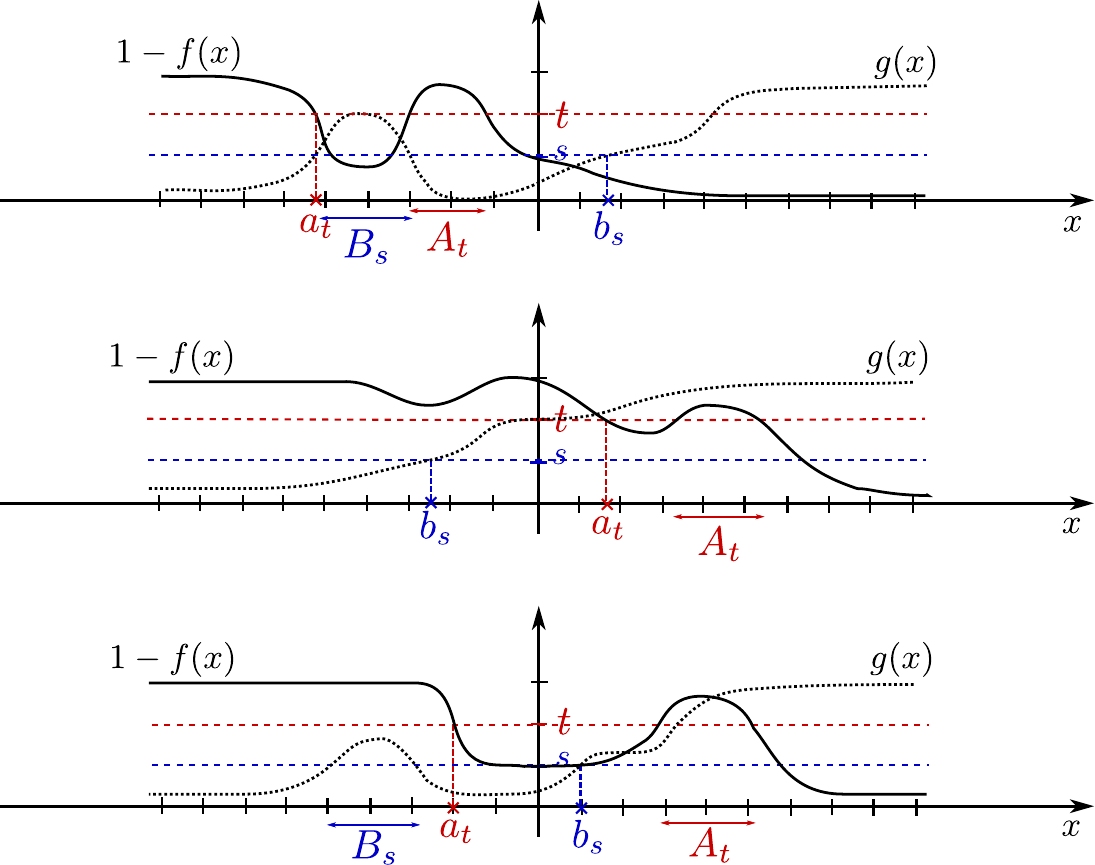}
\caption{Illustration of level sets used in the proof of Lemma~\ref{lem:elemrearr}. a) For $a_t+|A_t| < b_s-|B_s|$. The intersection on the right-hand side of~\eqref{eqn:keyineq} is empty and the inequality is trivial. b) For $a_t+|A_t| > b_s-|B_s|$ and $a_t \ge  b_s$ 
\eqref{eqn:keyineq} is an equality. c) For the last case $a_t+|A_t| > b_s-|B_s|$ and $a_t<b_t$ the inequality 
\eqref{eqn:keyineq} is non-trivial.}
\label{casesHL}
\end{figure}

\begin{lemma}\label{lem:elemrearr}
For any interpolating $f$ and $g$, we have
\[
\int_\reals {\mathrm d}x (1-f(x)) g(x)  \ge
\int_\reals {\mathrm d}x (1-\bar f(x)) \bar g(x)
\]
\end{lemma}
\begin{IEEEproof}If the left-hand side is infinite, then the result is immediate, so we assume that it is finite.

This result is very similar to the Hardy-Littlewood inequality for symmetric rearrangements.
We will, however, give a self-contained elementary proof.
The key inequality is the following which holds for all $t,s \in (0,1).$
\begin{align}
\begin{split}\label{eqn:keyineq}
&| \{x:1-f(x) >  t\} \cap \{x: g(x) >  s\} |
 \ge
| \{x:1-\bar f (x)>  t\} \cap \{x: \bar g(x) >  s\} |.
\end{split}
\end{align}
This gives the result since
\begin{align*}
\int_\reals {\mathrm d}x \, (1-f(x)) g(x) 
&=
\int_\reals {\mathrm d}x \int_0^1 {\mathrm d}t \, \indicator{1-f(x) >  t}(x)   \int_0^1 {\mathrm d}s \, \indicator{g(x) >  s}(x)  
\\
&=
\int_0^1\int_0^1 {\mathrm d}t \, {\mathrm d}s \, | \{x:1-f(x) >  t\} \cap \{x: g(x) >  s\} |.
\end{align*}

To see~\eqref{eqn:keyineq}, observe that, for $s,t\in (0,1),$ we have some maximal $a_t$ and minimal $b_s$ such that $\{x:1-f(x) >  t\} = (-\infty,a_t) \,\,\cup A_t$
and $\{x:g(x) >  s\} = (b_s,+\infty) \,\,\cup B_s$ where the unions are disjoint and
$|A_t|,|B_s| <\infty.$
If $a_t+|A_t| < b_s-|B_s|$ (see case a) in Fig.~\ref{casesHL}) then the right-hand side of~\eqref{eqn:keyineq} is $0$ 
and~\eqref{eqn:keyineq} is immediate, so we assume otherwise.  
If $a_t \ge b_s$ (see case b) in Fig.~\ref{casesHL}) then we trivially have equality in~\eqref{eqn:keyineq} so we also assume
$a_t < b_s.$ 
We now have case c) in Fig.~\ref{casesHL} and we obtain
\begin{align*}
 | \{x:1-\bar f (x)>  t\} \cap \{x: \bar g(x) >  s\} |
=& |\,(-\infty,a_t+|A_t|)\,\,\cap \,\, (b_s-|B_s|,+\infty) \,|
\\=& |A_t|+|B_s| - (b_s-a_t).
\end{align*}
Note that the last line is non-negative because we are {\it not} in the case  $a_t+|A_t| < b_s-|B_s|$.
Now, $A_t$ and $B_s$ can intersect only in the interval
$[a_t,b_s]$ so we have
\begin{align*}
| \{x:1- f (x)>  t\} \cap \{x: g(x) >  s\} |
=& |A_t|+|B_s| - |(A_t \cup B_s) \cap [a_t,b_s] |
\\ \ge & |A_t|+|B_s| - (b_s-a_t),
\end{align*}
and the lemma follows.
\end{IEEEproof}

\begin{lemma}\label{lem:elemsmthrearr}
For any interpolating $f$ and $g$, we have
\begin{align}\label{HL-style}
\int_\reals {\mathrm d}x (1-f^\smthker(x)) g(x)  \ge
\int_\reals {\mathrm d}x (1-\bar f^\smthker(x)) \bar g(x).
\end{align}
\end{lemma}
\begin{IEEEproof}
If the left-hand side is infinite, the inequality holds. Hence, we suppose it is finite. We have 
\begin{align*}
\int_\reals & {\mathrm d}x \, (1-f^\smthker(x)) g(x) 
= \int_\reals {\mathrm d}x \int_\reals {\mathrm d}y\,
w(y) (1 - f(x-y))g(x).
\end{align*}
Since the integrand is non-negative and the integral is finite, we can apply the Fubini theorem to rewrite
\begin{align}\label{Fubi}
\int_\reals & {\mathrm d}x \, (1-f^\smthker(x)) g(x) 
= \int_\reals {\mathrm d}y \, \smthker (y) \int_\reals \mathrm{d}x \, (1-f(x))g(x-y) .
\end{align}
Now, we apply Lemma~\ref{lem:elemrearr} to the functions $f$ and $g_y$, where $g_y(x) = g(x-y)$. Note that $g_y$ is simply 
a translated version of $g$, so its rearrangement is just obtained by the same translation of $\bar g$, i.e.,
$\widebar{g_y}(x) = \bar g(x-y)$. Thus,
\begin{align*}
\int_\reals \mathrm{d}x \, (1-f(x))g(x-y) \geq \int_\reals \mathrm{d}x \, (1-\bar f(x)) \bar g(x-y) .
\end{align*}
Multiplying by $w(y)$, integrating over $y$, and using~\eqref{Fubi}, we obtain~\eqref{HL-style}.

\end{IEEEproof}

We are now ready to prove a rearrangement inequality for $\mathcal{W}.$

\begin{proposition}[Monotonicity of Minimizers]\label{prop:mainRearrange}
Let $f$ and $g$ be profiles satisfying~\eqref{eqn:RearrLimits}
and let $\bar f$ and $\bar g$ be their respective increasing rearrangements.
Assume the PGC and that $C_\smthker < \infty,$ then we have
\begin{equation}\label{eqnIncRearrB}
\mathcal{W}(f,g) \geq \mathcal{W}(\bar f,\bar g).
\end{equation}
\end{proposition}

\begin{IEEEproof}
If the left-hand side of~\eqref{eqnIncRearrB} is infinite, then the result is immediate, so we assume that $\mathcal{W}(f,g)$
is finite.
Let us first assume that $\int_\reals \mathrm{d}x \, (1-f^\smthker (x))g(x) < \infty$ (in fact, we can assume the saturated case).
It then follows that
$L(f,g)$ in Equ.~\eqref{eqn:Lfg}
is finite.  Note that if $F:\mathds{R}\to [0,1]$ is monotone, then $\widebar{F\circ p} = F\circ \bar{p}$. Thus the increasing rearrangement of 
$\int_0^{f(x)} \mathrm{d}v\,h^{-1}_f(v)$  is equal to
$\int_0^{\bar f(x)} \mathrm{d}v\,h^{-1}_f(v)$ and similarly for the
term $\int_0^{g(x)} \mathrm{d}u\,(1-h^{-1}_g(u))$. 
We can now apply Lemma~\ref{lem:pqinvar} (with suitable scaling)
to conclude that $L(f,g)=L(\bar f, \bar g).$ 
For this case, the proposition now follows from Lemma ~\ref{lem:elemsmthrearr}.

Now, we consider the general case, where possibly $\int_\reals \mathrm{d}x \, (1-f^\smthker (x))g(x) = \infty$.
Due to Lemma~\ref{lem:klimW}, we have
\begin{equation}\label{eqn:lemRear1}
\mathcal{W}(f,g) =
\lim_{K\rightarrow \infty}
\mathcal{W}(\cut{f}{K},\cut{g}{K}).
\end{equation}
We remark that $\int_\reals  \mathrm{d}x \, (1-(\cut{f}{K}^\smthker (x))\cut{g}{K}(x) < \infty$ due to Lemma~\ref{lem:klimrearr}
 Equ.~\eqref{eqn:tailconv0}. Therefore, using the saturated case, we have already established above, we have
\begin{equation}\label{eqn:lemRear3}
\mathcal{W}(\cut{f}{K},\cut{g}{K}) \ge
W(\overline{ \cut{f}{K}}, \overline{ \cut{g}{K}}).
\end{equation}
Finally, it is easy to see that for any interpolating profile $f$, we have
$\overline{ \cut{f}{K}} \rightarrow \overline{f}$ pointwise.
By Lemma~\ref{lem:posfatou} Equ.~\eqref{eqn:Wfatou}, we  obtain
\begin{equation}\label{eqn:lemRear2}
\liminf_{K \rightarrow \infty}
W(\overline{ \cut{f}{K}}, \overline{ \cut{g}{K}}) \ge
W(\bar f, \bar g).
\end{equation}
Combining~\eqref{eqn:lemRear1},~\eqref{eqn:lemRear3}, and~\eqref{eqn:lemRear2} concludes the proof.
\end{IEEEproof}

Proposition~\ref{prop:mainRearrange} shows that minimizers $f$, $g$ of the 
functional $\mathcal{W}(f, g)$ can be found in the spaces of \emph{non-decreasing profiles}.
From now on, we therefore restrict the functional to those spaces.

\section{Existence of Minimizers}\label{sectionExistence}

The existence of a monotonic FP, which we will show is a minimizer of ${\mathcal W},$ is proved in \cite{KRU12}.
In this section, we give an alternate proof, under similar conditions, using the direct method of the calculus of variations
\cite{dacorogna}, as was done in \cite{ElKhatib}.
 
In the direct method of the calculus of variations, one constructs a minimizer as a limit point of a minimizing sequence.
Since ${\cal W}(f,g)$ is invariant under a common translation of $f$ and $g$, it is necessary to center the sequence in order to carry
out the method.  We can do this by translating $f$ and $g$ so that $\frac{1}{2}\in [f(0-),f(0+)].$  We call such a profile pair
{\em centered}.
 
\begin{proposition}\label{thm:existence}
Assume $C_\smthker < \infty$ and assume the SPGC is satisfied.
Then, there exists a monotonic non-decreasing  profile pair $(f(x),g(x))$ that minimizes ${\cal W}$
under the condition that $(f,g)$ has limit $(1,1)$ at $x=\infty$ and limit $(0,0)$ at $x=-\infty.$
\end{proposition}
\begin{IEEEproof}
We already remarked that we can adopt the alternative expression~\eqref{eqn:altWdefinition} for the 
potential functional, namely
\begin{align*}
{\cal W}(f,g) = \int_\reals {\mathrm d}x\, \tilde{I}_{f,g,\smthker}(x) ,
\end{align*}
where $\tilde{I}_{f,g,\smthker}(x) \geq 0$. Therefore ${\cal W}(f,g)$ is bounded from below so, 
by Proposition~\ref{prop:mainRearrange}, there exists a minimizing sequence $(f_i,g_i)$ of 
monotonic profiles satisfying the limit condition, i.e.,
$$
\lim_{i\to +\infty} {\cal W}(f_i,g_i) = \inf\mathcal{W}(f,g).
$$
Let us center the sequence so that $\frac{1}{2}\in [f_i(0-),f_i(0+)]$ for each $i.$
Interpreting $f_i$ and $g_i$ as cumulative probability distributions, our aim is to show the tightness 
of the sequence, i.e., that the transition of $f_i$ and $g_i$ from $\epsilon$ to $1-\epsilon$ must occur 
in a bounded region for all $i.$ 
 
Let $C$ be an arbitrary finite constant. Then, we claim that
that for any $\epsilon>0$ there exists $Z<\infty$ such that ${\cal W}(f,g) < C$
implies that  $f(x),g(x) > 1-\epsilon$ for $x>Z$ and $f(x),g(x) < \epsilon$ for $x < -Z,$ 
(assuming $f,g$ is a centered monotonic profile pair satisfying the limit conditions).

This claim completes the proof. Indeed we can then extract from $(f_i,g_i)$ a subsequence $(f_{i_k},g_{i_k})$ converging 
to a limit point $(f_*, g_*)$ which
necessarily satisfies the limit conditions, and by Fatou's lemma
\begin{align*}
 \int_\reals {\mathrm d}x\, \tilde{I}_{f_{*},g_{*},\smthker}(x) 
 \leq \liminf_{k\to +\infty} \int_\reals {\mathrm d}x\, \tilde{I}_{f_{i_k},g_{i_k},\smthker}(x),
\end{align*}
so $\mathcal{W}(f_*, g_*) \leq \inf\mathcal{W}(f,g)$ and $f_*, g_*$ is a monotone minimizing pair for the potential functional. 
 
Now we prove the claim. 
Since, by Lemma~\ref{lem:Cbound}, we have
\begin{align*}
\int_\reals |f^\smthker(x) -f(x)| \, dx \le  C_\smthker,
\end{align*}
we see that  ${\cal W}(f,g) < C$ implies that
\begin{align*}
\int_\reals\phi & (h_f,h_g;g(x),f(x)) \, dx 
= \mathcal{W}(f,g) - \int_\reals \mathrm{d}x\, (f(x)-f^w(x))g(x)
\le C + C_\smthker\,.
\end{align*}
By the strictly positive gap condition, there exists $\eta >0$ such that $\phi(h_f,h_g; g(x),f(x)) > \eta$
unless we have either $f(x),g(x) < \epsilon$ or $f(x),g(x) > 1-\epsilon.$ Let $x_{+}$ be the least $x$ such that $f(x), g(x) > 1-\epsilon$.
Then, 
\begin{align*}
C+C_{\smthker} \ge \int_0^{x_+}\phi & (h_f,h_g;g(x),f(x)) \, dx > \eta x_{+}, 
\end{align*}
and $x_{+} < (C+C_{\smthker})/\eta$.
Thus for each $i$, we have
$f_i(x),g_i(x) > 1-\epsilon$ for $x > (C+C_{\smthker})/\eta$. Similarly, we have for each $i$ we have
$f_i(x),g_i(x) < \epsilon$ for $x < -(C+C_{\smthker})/\eta.$
\end{IEEEproof}

\section{Displacement Convexity}\label{sectionDC}

A generic functional $\mathcal{F}(p)$ on a space $\mathcal{X}$ (of profiles, say) is said to be convex in the usual sense if, for any pair $p_0,p_1\in\mathcal{X}$, and for all $\lambda\in[0,1]$, and for the {\it linear interpolation} $(1-\lambda)p_0+\lambda p_1$ of the profiles, the inequality
$\mathcal{F}((1-\lambda)p_0+\lambda p_1)\leq (1-\lambda)\mathcal{F}(p_0)+\lambda \mathcal{F}(p_1)$ holds.
Displacement convexity, on the other hand, is defined as convexity under an alternative interpolation called {\it displacement interpolation}. The usual setting for displacement convexity is a space of probability measures. For measures over the real line, one can conveniently define the displacement interpolation in terms of the cdf's associated to the measures. This is the simplest setting and the one that we adopt here.

We think of the increasing profiles $p$ as right-continuous cdf's of some underlying measures $\mathrm{d}p$
over the real line. As already stated, the inverse $p^{-1}(u)$ defined almost everywhere and with left and right limits $p^{-1}(u_{-})$ and $p^{-1}(u_{+})$, respectively, 
are uniquely defined. However, at this point, it is useful to settle on the right-continuous inverse which is defined for all $u\in (0,1)$, 
namely $p^{-1}(u) = \inf\{x\mid p(x) > u\}$.

Consider two profiles $p_0$ and $p_1$, and assume $p_0$ is 
continuous.  We can define a map $T_p : \reals \to \reals$ as
\begin{equation}
T_p(x) = p_1^{-1}(p_0(x)).
\end{equation}
The map $T_p$ can be seen as a pushforward map for measures from $\mathrm{d}p_0$ to $\mathrm{d}p_1$. 
This is expressed as $\mathrm{d}p_1=T_{p}\# \mathrm{d}p_0$ which means 
\begin{align*}
 \int \mathrm{d}p_1(x)\, h(x) = \int \mathrm{d}p_0(x)\, h(T_p(x))
\end{align*}
for any function $h$ such that the integral is well-defined. 
Then, denoting by $id$ the identity map, the interpolant $p_\lambda(\cdot)$ is the cdf of the measure defined by
\begin{align*}	
\mathrm{d}p_\lambda=((1-\lambda)id+\lambda T_p)\# \mathrm{d}p_0.
\end{align*}
We have
\[
\int \mathrm{d}p_\lambda (x) h(x)
=
\int \mathrm{d}p_0 (x) h((1-\lambda) x+\lambda T_p(x)),
\]
whenever the integral is defined.  In particular, if $h$ is convex, then this shows
convexity in $\lambda$ of the integral due to the following,
\begin{align*}
\int \mathrm{d} p_0 (x) h((1-\lambda)(x)+\lambda T_p(x))
&\leq (1-\lambda)\int \mathrm{d} p_0 (x) h(x)+\lambda \int \mathrm{d}p_0 (x) h(T_p(x))\\
& = (1-\lambda)\int \mathrm{d} p_0 (x) h(x)+\lambda \int \mathrm{d} p_1 (x) h(x).
\end{align*}

\begin{figure}
\centering
\includegraphics[draft=false,scale=0.6]{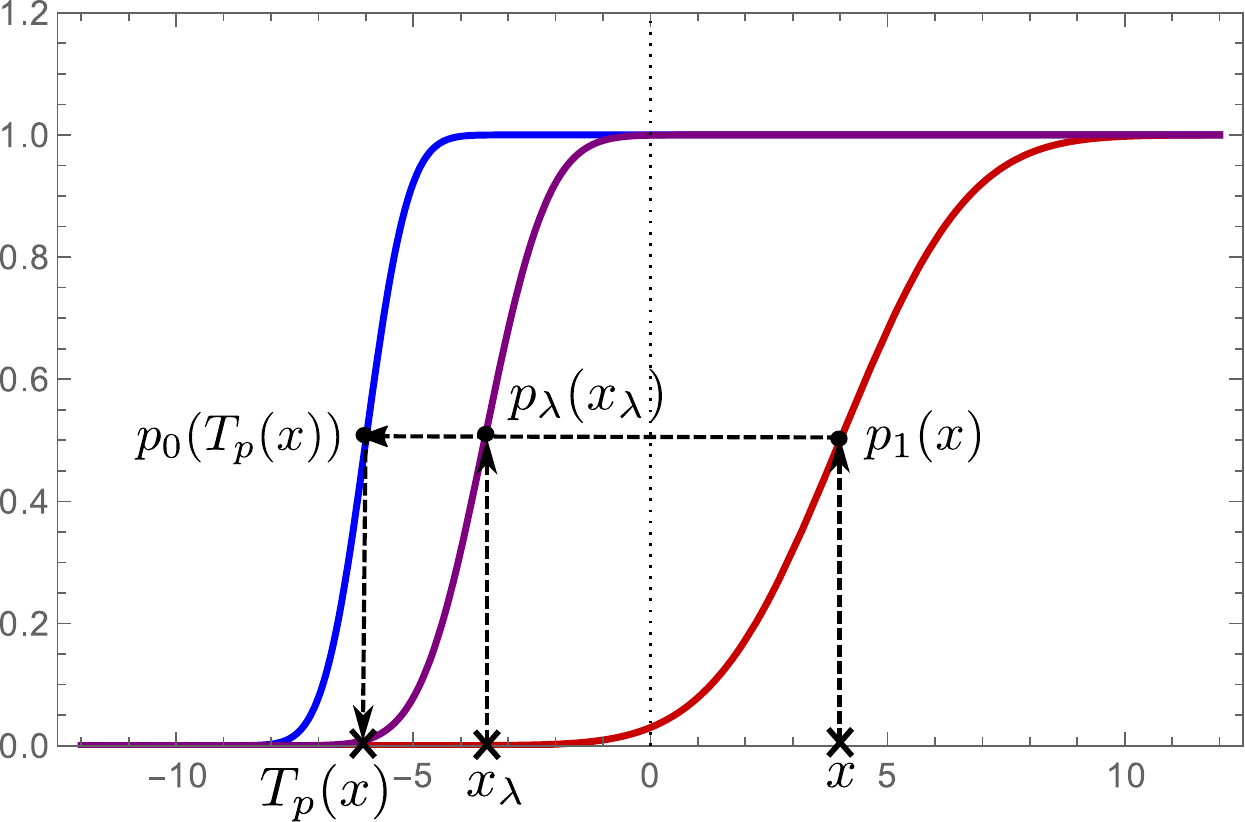}
\caption{Monotonic profiles $p_0$ and $p_1$, the map $T_p(x)$, $x_{p,\lambda} = (1-\lambda) x + \lambda T_p(x)$, and 
the interpolant $p_\lambda$.
Here, $\lambda=1/4$.} 
\label{displacement-picture}
\end{figure}

The graphical construction of the interpolant $p_\lambda$ is illustrated in Fig.~\ref{displacement-picture}.
Graphically, $T_p$ finds the position $x^\prime$ on the $x$-axis so that $p_1(x^\prime)=p_0(x)$ for some given $x$. 
Consider the linear interpolation between points on $\mathds{R}$,
$
x_{p,\lambda} = (1-\lambda)x+\lambda T_p(x).
$
The displacement interpolant $p_\lambda$ is defined so that the following equality 
holds for all $\lambda\in [0,1]$ 
%and almost all $x\in \mathds{R}$
$
p_\lambda(x_{p,\lambda}) = p_0(x).
$

In the case where $p_0$ is discontinuous, we have to be more careful in the definition.
At points of discontinuity of $p_0$, the map $T_p(x)$ should not be single-valued.
Since we work in one dimension, this issue is easily circumvented and 
we can in general define $p_\lambda$ via its inverse as
\begin{equation}\label{eqn:fInvLambda}
p^{-1}_\lambda(u) =(1-\lambda)p_0^{-1}(u)+\lambda p_1^{-1}(u),
\end{equation}
and $p_\lambda(x) = \inf\{ u \mid p_\lambda^{-1}(u) > x\}$ (which is right continuous). 
% and at discontinuities of $p^{-1}_\lambda$ 
% invoke monotonicity to define $p_\lambda(x)$ for all $x$ (in the continuous case this is equivalent to 
%~\eqref{eqnFlambda}). 
Correspondingly, if $p$ is an interpolating increasing profile then, under appropriate regularity of $h,$ we can write
\[
\int_\reals \mathrm{d} p(x) h(x) =\int_0^1 \mathrm{d}u\, h(p^{-1}(u))\,
\]
and we have
\[
\int \mathrm{d} p_\lambda (x) h(x)
=
\int_0^1 \mathrm{d}u\, h((1-\lambda)p_0^{-1}(u)+\lambda p_1^{-1}(u))\,.
\]
With this in mind, we will continue to use the notation $T_p(x)$ when the above interpretation should be understood.

In the remainder of this work, we consider two pairs of interpolating profiles $(f_0,f_1)$ and $(g_0, g_1)$ and consider the corresponding interpolants $f_\lambda$ and $g_\lambda$. 

We now state one of the main results of this paper.
\begin{proposition}\label{propConv}
Assume the PGC and  $C_\smthker < \infty.$
Then, the potential $\mathcal{W}(f,g)$ is displacement convex; that is, for all $\lambda\in [0,1]$,
\begin{align}\label{eqn:propConv}
\mathcal{W}(f_\lambda,g_\lambda) \leq (1-\lambda) \mathcal{W}(f_0,g_0) + \lambda \mathcal{W}(f_1,g_1).
\end{align}
\end{proposition}

We first show that it is sufficient to prove the proposition 
under the assumption that
$(f_0,g_0)$ and $(f_1,g_1)$ are saturated.
We recall that by Lemma~\ref{lem:klimW},
for any monotonic interpolating pair $f,g$ we have 
\begin{equation}\label{lem:KWconverge}
\lim_{K\rightarrow \infty}
\mathcal{W}(\cut{f}{K},\cut{g}{K}) =
\mathcal{W}(f,g)\,.
\end{equation}
Given any monotonic interpolating pairs $(f_0,g_0), (f_1,g_1)$, 
let
 $f_{K,\lambda}$ denote the displacement interpolant of 
$\cut{f_0}{K}$ and $\cut{f_1}{K}.$
It is easy to see that 
$f_{K,\lambda}$ converges pointwise to $f_\lambda$ when $K\to +\infty$.
By Lemma~\ref{lem:posfatou} Equ.~\eqref{eqn:Wfatou} we therefore have
\begin{align}\label{secondinequ}
\liminf_{K\rightarrow \infty} \mathcal{W}(f_{K,\lambda},g_{K,\lambda}) \ge
\mathcal{W}(f_{\lambda},g_{\lambda})\,.
\end{align}
In view of~\eqref{lem:KWconverge} and~\eqref{secondinequ}, we see that~\eqref{eqn:propConv} follows from
\begin{align}
 \mathcal{W}(&f_{K,\lambda},g_{K,\lambda})
 \leq (1-\lambda) \mathcal{W}(\cut{f_0}{K},\cut{g_0}{K}) + \lambda \mathcal{W}(\cut{f_1}{K},\cut{g_1}{K}).
 \label{conv-displ-sat}
\end{align}
which is the saturated case of~\eqref{eqn:propConv}.
For the remainder of the section, we therefore assume the saturated case, and prove~\eqref{conv-displ-sat}.

If $f$ and $g$ are saturated then we have
$$
\int_{\reals} {\mathrm d}x \, (1-f^\smthker(x))g (x)< \infty\,.
$$
Indeed, 
$$
(1-f^\smthker(x))g (x) = (1-f(x))g(x) + (f(x) - f^w(x))g(x),
$$  
and the first term is integrable for saturated
profiles $f, g$, and the second term is also integrable because of  Lemma~\ref{lem:Cbound} (note that $f^w$ is not necessarily saturated).
This is the critical requirement since,
by integrating by parts, we obtain
\begin{align}\label{eqn:winnerform}
\int_{\reals} {\mathrm d}x \, (1-f^\smthker(x))g (x) =
\iint_{\reals^2} \mathrm{d} f(x) V(x-y)\mathrm{d}g(y).
\end{align}
The full derivation of this identity reads
\begin{align}
\int_{\reals} &{\mathrm d}x \, (1-f^\smthker(x))g(x)
=\iint_{\reals^2} {\mathrm d}x \,{\mathrm d}y \, (1-f(x))\smthker(x-y)g(y) \nonumber
\\
&=
\iint_{\reals^2} \mathrm{d}f(x) \Omega(x-y) g(y)\,{\mathrm d}y 
=
\iint_{\reals^2} \mathrm{d}f(x) V(x-y)\mathrm{d}g(y),\,\nonumber
\end{align}
where we have used the fact the $V(x)$ is well-defined. 

The identity~\eqref{eqn:winnerform} leads to the following key result:

\begin{lemma}\label{lem:convCrossterm}
Let $(f_0,g_0)$ and $(f_1,g_1)$ be saturated, then
\begin{align*}
\int_{\reals} {\mathrm d}x \, (1-f_\lambda^\smthker(x))g_\lambda (x) 
\end{align*}
is a convex function of $\lambda.$
\end{lemma}

\begin{IEEEproof}
Since $f_\lambda$ and $g_\lambda$ are saturated we have 
by~\eqref{eqn:winnerform} 
%and Lemma~\ref{lem:Cbound} 
that
\begin{align*}
&\int_{\reals} (1-f_\lambda^\smthker(x))g_\lambda (x)\,{\mathrm d}x
 =
\iint_{\reals^2} \mathrm{d}\!f_\lambda (x) V(x-y)\mathrm{d}g_\lambda(y)
\\
 & =
\iint_{\reals^2} \mathrm{d}\!f_0 (x) V((1-\lambda)(x-y)
+\lambda(T_f(x)-T_g(y)))\mathrm{d}g_0(y).
\end{align*}
This is convex in $\lambda$ because the kernel $V$ is convex (see~\ref{lemVconvex}). 
\end{IEEEproof}

\begin{lemma}\label{lem:Llinear}
For any saturated pairs $(f_0,g_0),(f_1,g_1),$ the functional $L(f_\lambda,g_\lambda)$ is affine in $\lambda.$ 
\end{lemma}
\begin{IEEEproof}
We will show that $L(f_\lambda,g_\lambda)-L(f_0,g_0)$ is linear in $\lambda$. We start by the considering the first term of this difference.
Using the layer cake representation and the monotonicity of the functions, we have
\begin{align*}
\int_\reals \mathrm{d}x \Big(\int_0^{f_\lambda(x)}\mathrm{d}u \, h_f^{-1} ( u )
-\int_0^{f_0(x)} \mathrm{d}u \, h_f^{-1} ( u ) \Big) 
&=\int_\reals \mathrm{d}x\,  \int_{f_0(x)}^{f_\lambda(x)}  \mathrm{d}u\, h_f^{-1}(u)
\nonumber \\ &
= \int_0^1 \mathrm{d}u \, h_f^{-1}(u) (f_0^{-1}(u)-f_\lambda^{-1}(u)) .
\end{align*}
Using~\eqref{eqn:fInvLambda} we can write write this as
\begin{align*}
\lambda \int_0^1 \mathrm{d}u \, h_f^{-1}(u) (f_0^{-1}(u)-f_1^{-1}(u)),
\end{align*}
which is evidently linear in $\lambda$. 
Similarly for the second term in the difference $L(f_\lambda,g_\lambda)-L(f_0,g_0)$, we obtain
%Applying this to the various terms we obtain
%\begin{align*}
%L(f_\lambda,g_\lambda)-&L(f_0,g_0) \\
%&=\lambda\Biggl(
%\int_0^1 \mathrm{d}u  (g_1^{-1}(u)-g_0^{-1}(u))   (h_g^{-1}(u)-1)
%\\
%&+
%\int_0^1 \mathrm{d}u  (f_1^{-1}(u)-f_0^{-1}(u))   h_f^{-1}(u)
%\Biggr)\,.
%\end{align*}
\begin{align*}
\int_\reals &\mathrm{d}x \Big(\int_0^{g_0(x)}\mathrm{d}u \, (1-h_g^{-1} ( u ))
-\int_0^{g_\lambda(x)} \mathrm{d}u \, (1-h_g^{-1} ( u )) \Big) \nonumber \\
&=\lambda \int_0^1 \mathrm{d}u \, (1-h_g^{-1}(u)) (g_1^{-1}(u)-g_0^{-1}(u)) \,.
\end{align*}
\end{IEEEproof}

We are now ready to prove the main result of this section.

\begin{IEEEproof}[Proof of Proposition~\ref{propConv}]
If $\mathcal{W}(f_0,g_0)=+\infty$ or $\mathcal{W}(f_1,g_1)=+\infty$, then the result is immediate,
so we assume both are finite.
As argued above, 
we can assume that all functions are saturated.
We rewrite the potential in~\eqref{eqnPotBothFts} as follows
\begin{align}\label{eqnPotDC}
\mathcal{W}&(f_\lambda ,g_\lambda )=L(f_\lambda ,g_\lambda )
+\int_{\mathds{R}}{\mathrm d}x\, (1-f_\lambda^\smthker(x))g_\lambda(x).
\end{align}
By Lemma~\ref{lem:Llinear}, the functional $L(f_\lambda,g_\lambda)$ is affine and hence convex in $\lambda.$
The second term was shown to be convex in Lemma~\ref{lem:convCrossterm}.
\end{IEEEproof}

\section{Fixed Points and Minimizers}\label{sectionCFP}

The main goal of this section is to prove Proposition~\ref{prop:CFPmin}, which states that a pair of monotonic 
profiles minimizes $\mathcal{W}$ if and only if it is a ``consistent" fixed point (CFP). It will be helpful to start with a 
preliminary discussion motivating the definition of CFP. 

We already remarked that $\phi(h_f, h_g;u,v)$ is convex in $v$ for fixed $u$ and minimized (over $v$) by setting $v=h_f(u)$, and similarly for $u$ and $v$ interchanged.  From~\eqref{eqn:Idef}, a similar argument shows that
$I_{f,g,\smthker}(x) \ge I_{f,h_g \circ f^\smthker,\smthker}(x)$
and $I_{f,g,\smthker}(x) \ge I_{h_f \circ g^\smthker,g,\smthker}(x)$
so that
\begin{align*}
{\cal W}(f,g) \ge {\cal W}(f,h_g \circ f^\smthker) {\rm ~and~} 
{\cal W}(f,g) \ge {\cal W}(h_f \circ g^\smthker,g).
\end{align*}
Under some conditions, we can have 
${\cal W}(f,g) = {\cal W}(f,h_g \circ f^\smthker)$ even though
it is not the case that $g = h_g \circ f^\smthker$ almost everywhere.
This can happen, in particular, if $h_g$ is discontinuous and the pair $h_f,h_g$ does not 
satisfy the strictly positive gap condition.

One of the main analytical tools used in \cite{KRU12} was the construction of $h_f$ and $h_g$ given
$f,g,$ and $\smthker$ so that $f,g$ form a ``consistent" interpolating fixed point.  Note that, from an interpolating fixed point, we can recover
the graph $(u,h_f(u)), u\in[0,1]$ of $h_f$ as the parametric curve $(g^\smthker(x),f(x))$ as $x\in (-\infty,+\infty).$
Given interpolating $f$ and $g$, we denote the $h_f$ so obtained as $h_{[f,g^\smthker]}$
(see \cite{KRU12}  for more detail.)
The update function $h_{[f,g^\smthker]}$ is uniquely determined at points of continuity but may not be uniquely determined
at points of discontinuity.  In particular, if $g^\smthker$ is constant over some open interval $I$ where $f$ is increasing then
$h_{[f,g^\smthker]}$ has a discontinuity at that value of $g^\smthker(I)$ and we see that we cannot have
$f = h_{[f,g^\smthker]} \circ g^\smthker$ almost everywhere.  Nevertheless, it is the case that
$f(x) \in  [h_{[f,g^\smthker]}(g^\smthker (x)-),h_{[f,g^\smthker]}(g^\smthker (x)+)]$  for all $x$
and in this sense it satisfies the DE equation.
In \cite{KRU12}, the notation 
\begin{align*}
f \veq h_f \circ g^\smthker
\end{align*}
was used to capture this case.\footnote{More precisely, if $h$ plays the role of $h_f$, $h_g$ we denote by $v \veq h(u)$ when $v=h(u)$ at points of continuity of $h$ and $v\in [h(u-), h(u+)]$ at points of discontinuity of $h$.}  This motivates the following definition:

\begin{definition} 
We say that an interpolating pair $f,g$ of profiles is a {\it consistent fixed point} (CFP) if 
$f \veq h_f \circ g^\smthker$
and $g \veq h_g \circ f^\smthker$. Recall that $f,g$ is a fixed point (FP) if $f = h_f \circ g^\smthker$ and 
$g = h_g \circ f^\smthker$ almost everywhere, i.e., up to a set of measure zero.
\end{definition}

\begin{proposition}\label{prop:CFPmin}
Let $f,g$ be monotonic and interpolating. Then 
$\mathcal{W}(f,g)$ is minimal - in the sense 
$\mathcal{W}(f,g) \le \mathcal{W}(f',g')$
for any monotonic interpolating $f',g'$ - if and only if
$f,g$ is a CFP.
\end{proposition}

\begin{IEEEproof}
If $f,g$ is not a CFP then either
$\mathcal{W}(f,g) = \infty$ in which case the pair cannot be minimal, or we have either
${\cal W}(f,g) > {\cal W}(f,h_g \circ f^\smthker)$
or
${\cal W}(f,g) > {\cal W}(h_f \circ g^\smthker,g)$,
which shows that $\mathcal{W}(f,g)$ is not minimal.

To prove the converse, assume $f_0,g_0$ is a CFP.
The proof proceeds by contradiction.
Hence, we suppose there exists interpolating $f_1,g_1$ with
$\mathcal{W}(f_0,g_0) > \mathcal{W}(f_1,g_1)$ and we shall deduce a contradiction.
By Lemma~\ref{lem:klimW}, we can assume that $f_1$ and $g_1$ are saturated.

We will show that we may also take $f_0,g_0$ to be saturated.
Define
\begin{align*}
h_{f_0}^K  &= h_{[\cut{f_0}{K},\cut{g_0}{K}^{\smthker}]}, \quad
h_{g_0}^K  = h_{[\cut{g_0}{K},\cut{f_0}{K}^{\smthker}]}
\end{align*}
so $\cut{f_0}{K},\cut{g_0}{K}$ is a CFP for $h_{f_0}^K,h_{g_0}^K.$
Since $f_1,g_1$ are saturated, it follows easily that
\[
\lim_{K\rightarrow\infty} \mathcal{W}(h_{f_0}^K,h_{g_0}^K;f_1,g_1)
=\mathcal{W}(h_{f_0},h_{g_0};f_1,g_1) = \mathcal{W}(f_1,g_1),
\]
and by Lemma~\ref{lem:minlimit} we have
\[
\lim_{K\rightarrow\infty} \mathcal{W}(h_{f_0}^K,h_{g_0}^K;\cut{f_0}{K},\cut{g_0}{K})
= \mathcal{W}(f_0,g_0),
\]
and we see that we can assume $f_0,g_0$ are saturated.

Since $f_0,g_0$ is a CFP
it follows that $\mathcal{W}(f_0,g_0) \le\mathcal{W}(f',g_0)$
and $\mathcal{W}(f_0,g_0) \le\mathcal{W}(f_0,g')$ for all interpolating $f'$ and $g'.$
Hence, we now have
\begin{align}
\mathcal{W}(f_\lambda,g_\lambda)
-\mathcal{W}(f_0,g_0) 
& \ge
\mathcal{W}(f_\lambda,g_\lambda)
-\mathcal{W}(f_\lambda,g_0)
-\mathcal{W}(f_0,g_\lambda)
+\mathcal{W}(f_0,g_0)\nonumber
\\
& = -\int_\reals (f^\smthker_\lambda(x)-f^\smthker_0(x))(g_\lambda(x)-g_0(x))\, {\mathrm d}x\nonumber
\\& \ge
- C\lambda^2 \label{eqn:quadbnd},
\end{align}
where $C$ is some positive constant.  
The last step follows from $| f^\smthker_\lambda(x)-f^\smthker_0(x)| \le C_1 \lambda$ and
$\int_\reals {\mathrm d}x \, | g_\lambda(x)-g_0(x) | \,  \le C_2 \lambda$ for some positive constants $C_1$ and $C_2,$
which follows from the saturation of $f_0,g_0$ and $f_1,g_1.$
By Proposition~\ref{propConv}, we have
\begin{align}\label{eqn:linbnd}
\mathcal{W}(f_\lambda,g_\lambda)
-\mathcal{W}(f_0,g_0)
\le
\lambda( \mathcal{W}(f_1,g_1)
-\mathcal{W}(f_0,g_0)).
\end{align}
Because of the assumption on $f_1, g_1$ the right-hand side of~\eqref{eqn:linbnd} is strictly negative.
Thus~\eqref{eqn:quadbnd} and~\eqref{eqn:linbnd} contradict each other for $\lambda$ sufficiently small.  
We conclude that no such $f_1,g_1$ can exist.
\end{IEEEproof}

We conclude this section in Lemma~\ref{lem:kappadef} with a pleasing expression  for $\mathcal{W}(f, g)$  when $f,g$ is a monotonic minimizer, equivalently a CFP. 
To obtain the expression, and for further application, we require a result concerning the following functional from \cite{KRU12},
\begin{align}
\begin{split}\label{eqn:altPhidef}
\altPhiSI&(\smthker; f,g;x_1,x_2) 
= \iint \mathrm{d}g(y) \mathrm{d}f(x) (\ind_{C_1}\Omega (x-y)
+ \ind_{C_2}\Omega (y-x))
\end{split}
\end{align}
where
\begin{align*}
C_1 = C_1(x_1,x_2) = \{ (x,y): x\le x_2, y>x_1 \}, \quad 
C_2 = C_2(x_1,x_2) = \{ (x,y): x > x_2, y\le x_1\}.
\end{align*}
Note that $\altPhiSI$ is non-negative; this is closely related to the positive gap condition.
One of the main results in \cite{KRU12} (Lemma 9) is the following (this result is used in Section~\ref{sectionUnicity}).
\begin{lemma}\label{lem:spatialintegration}
Let $f,g$ be a CFP for~\eqref{eqn:DE}, then
\begin{equation}\label{eqn:spatialintegration}
\altPhiSI(\smthker; f,g;x_1,x_2) 
=
\altPhi(h_f,h_g;g(x_1+),f(x_2+)).
\end{equation}
\end{lemma}
It turns out for our application that we only require the case 
$x_1=x_2$ and in this case the right-hand side of~\eqref{eqn:altPhidef}  simplifies,
at least at points of continuity of $f$ and $g,$ to
\begin{align}
\iint \mathrm{d}f(x) \mathrm{d}g(y)\, \mathbb{1}_{\{(x-x_1)(y-x_1)\le 0\}} \Omega(-|x-y|).
\label{eqn:altphiform}
\end{align}

\begin{lemma}\label{lem:kappadef}
If $f,g$ is a CFP then 
\begin{align*}
 \mathcal{W}(f,g)
= \iint_{\reals^2} \mathrm{d}f(x) \mathrm{d}g(y) \kappa(x-y),
\end{align*}
where 
\[
\kappa (x) := V(x) - x\Omega(x) = -\int_{-\infty}^x y \smthker(y) \, {\mathrm d}y.
\]
Note that $\kappa$ is a non-negative even function that tends to $0$ at
$\pm \infty$ (recall $w$ is an odd function). 

\end{lemma}

\begin{IEEEproof}
By Lemma~\ref{lem:minlimit}, it is enough to prove this for the saturated case.
For the saturated case, we can integrate by parts to obtain
\begin{align*}
\int_{\reals}(1-f(x))g (x)\,{\mathrm d}x
=
\iint_{\reals^2} \mathrm{d}\!f(x) \indicator{x-y\ge 0} (x-y) \mathrm{d}g(y)\,.
\end{align*}
From Lemma~\ref{lem:spatialintegration} and~\eqref{eqn:altPhidef} (or~\eqref{eqn:altphiform}) , we have
\begin{align}
\int_{\reals} & {\mathrm d}x \, \phi(h_f,h_g;g(x),f(x))
= \iint_{\reals^2} \mathrm{d} f(x)  |x-y|\Omega(-|x-y|) \mathrm{d}g(y)\,.
\end{align}
Combining these two equations we obtain
\begin{align*}
L(f,g)&
=\iint_{\reals^2} \mathrm{d}f(x) ( |x-y|\Omega(-|x-y|) - \indicator{x-y\ge 0} (x-y))\mathrm{d}g(y)
\\ &=
- \iint_{\reals^2} \mathrm{d}f(x)  (x-y)( \indicator{x-y\ge 0} -\Omega(-|x-y|))\mathrm{d}g(y)\,
\\ &=
- \iint_{\reals^2} \mathrm{d}f(x)  (x-y)\Omega(x-y)\mathrm{d}g(y)\,.
\end{align*}
Adding this to~\eqref{eqn:winnerform} yields the result by
the definition of $L(f,g)$ given in~\eqref{eqn:Lfg}.
\end{IEEEproof}

\section{Unicity of Minimizer}\label{sectionUnicity}

The existence of increasing interpolation solutions to~\eqref{eqn:DE} was established in \cite{KRU12} under the assumption of 
the strictly positive gap condition and assuming that $w$ is strictly positive on an interval $(-W,W),$ $W \le +\infty$
and $0$ off of $[-W,W].$  (We shall refer to this as the {\em interval support condition}.)
It was also shown in \cite{KRU12} that existence of such
a fixed point implies the positive gap condition and, by example, it was shown that if
$A(h_f, h_g;u) = 0$ for some $u \in (0,1)$, then there may be an infinite family
of fixed point solutions that are not equivalent under translation.
In this section we use displacement convexity to show that the solution whose existence
was proved in \cite{KRU12} under the strictly positive gap condition is unique up to displacement.  
%The proof relies on the potential function formulation.

It follows from Proposition~\ref{prop:CFPmin} that all interpolating minimizers have the same potential and that they are all CFPs.
By Proposition~\ref{propConv}, we see that if $f_0,g_0$ and $f_1,g_1$ are both monotonic interpolating CFPS then
$f_\lambda,g_\lambda$ is a CFP for all $\lambda \in [0,1].$
Displacement convexity can therefore not be strict in this case.
The aim of the proof is to show that the strictly positive gap condition then leads to the conclusion that 
all CFPs are equal up to translation.

Given $f_0,g_0$ and $f_1,g_1$, we define
\[
D(u,v) =(f_1^{-1}(v)- g_1^{-1}(u)) - (f_0^{-1}(v)- g_0^{-1}(u)). 
\]
  
\begin{lemma}\label{lem:displaceflat}
Let $f_0,g_0$ and $f_1,g_1$ be CFPs and assume the interval support condition.  Then, for all $\lambda \in [0,1]$, 
we have
\begin{align*}
\mu
\Bigl\{
(u,v) : &\,| f_\lambda^{-1}(v)-g_\lambda^{-1}(u)| < W,
D(u,v) \neq 0,\, \phi(u,v)\neq 0
\Bigr\}
=0,
\end{align*}
where $\mu$ denotes 2-d Lebesgue measure.
\end{lemma}
\begin{IEEEproof}
We assume throughout that
$f_0,g_0$ and $f_1,g_1$ are CFPs. Formally, we have
%Since $L(u,v)$  is linear in $\lambda$ we have, formally, 
\begin{align*}
&\frac{\mathrm{d}^2}{\mathrm{d}\lambda^2} {\cal W}(f_\lambda,g_\lambda)
=
\iint_{[0,1]^2} \mathrm{d}u \mathrm{d}v D(u,v)^2 \smthker (f_\lambda^{-1}(v)-g_\lambda^{-1}(u)).
\end{align*}
The formula is derived in Appendix~\ref{app:second-derivative} for saturated profiles. 
Note that the integrand is always non-negative so the integral is well-defined, although it may take the value $+\infty.$
We claim that
\begin{align}
\int_0^1 \mathrm{d}\lambda \iint_{{[0,1]^2}} \mathrm{d}u \mathrm{d}v D(u,v)^2 \smthker (f_\lambda^{-1}(v)-g_\lambda^{-1}(u)) = 0\,.
\label{eqn:claimedformula}
\end{align}
Assume that the claim is false.  Then there exists a set $A \subset {[0,1]^2}$ on which $f_0^{-1},g_0^{-1},f_1^{-1},g_1^{-1}$  are all  bounded such that
\[
\int_0^1 \mathrm{d}\lambda \iint_{{A}} \mathrm{d}u \mathrm{d}v D(u,v)^2 \smthker (f_\lambda^{-1}(v)-g_\lambda^{-1}(u)) =\eta > 0.
\]
In the saturated case, it is easy to see that ${\cal W}(f_\lambda,g_\lambda)$ is absolutely continuous and
so is $\frac{\mathrm{d}}{\mathrm{d}\lambda} {\cal W}(f_\lambda,g_\lambda).$
It now follows that for all $K$ large enough, we have
\begin{align*}
&\int_{[0,1]} \mathrm{d}\lambda \frac{\mathrm{d}^2}{\mathrm{d}\lambda^2} {\cal W}(f_{K,\lambda},g_{K,\lambda})
\ge \eta,
\end{align*}
and therefore, using the convexity of ${\cal W}(f_{K,\lambda},g_{K,\lambda})$ with respect to $\lambda$,
% \begin{align*}
% {\cal W}(\cut{f_1}{K},\cut{g_1}{K}) - {\cal W}({f_{K,\frac{1}{2}}},{g_{K,\frac{1}{2}}}) 
%  \leq \frac{1}{2}\frac{d}{d\lambda}{\cal W}(f_{K,\lambda},g_{K,\lambda})\vert_{\lambda=1}
%  \\
% {\cal W}(\cut{f_0}{K},\cut{g_0}{K}) - {\cal W}({f_{K,\frac{1}{2}}},{g_{K,\frac{1}{2}}}) 
% \leq - \frac{1}{2}\frac{d}{d\lambda}{\cal W}(f_{K,\lambda},g_{K,\lambda})\vert_{\lambda=0}\, ,
% \end{align*}
we deduce that 
there is  a positive constant $\gamma$ such that, for all $K$ large enough, we have
\[
{\cal W}(\cut{f_0}{K},\cut{g_0}{K})
+{\cal W}(\cut{f_1}{K},\cut{g_1}{K})
-2 {\cal W}({f_{K,\frac{1}{2}}},{g_{K,\frac{1}{2}}})
> \gamma\,.
\]
Applying Lemma~\ref{lem:klimW} and Lemma~\ref{lem:posfatou}\eqref{eqn:Wfatou}, and noting that
$f_{K, \lambda},g_{K, \lambda}$ converges pointwise to $f_{\lambda},g_{\lambda}$ yields 
\[
{\cal W}(f_0,g_0)
+{\cal W}(f_1,g_1)
-2 {\cal W}({f_{\frac{1}{2}}},{g_{\frac{1}{2}}})
\ge \gamma,
\]
which contradicts Proposition~\ref{prop:CFPmin}, thereby establishing the claim.  Note that the claim gives the
desired result except perhaps on a set of $\lambda$ of measure $0.$

Now assume that for some $\lambda \in [0,1]$ 
we have
\begin{align*}
\mu
\Bigl\{
(u,v) : &\,| f_\lambda^{-1}(v)-g_\lambda^{-1}(u)| < W,
\, D(u,v) \neq 0,\, \phi(u,v)>0
\Bigr\}
>0.
\end{align*}
By the continuity ($f_{\lambda}^{-1}$ and $g_{\lambda}^{-1}$ are continuous off of at most a countable set) and inner-regularity of Lebesgue measure,
there exists a closed set $A \in (0,1)^2$ of positive measure and a constant $\eta>0$ such that
for all $(u,v)\in A$ we have
$| f_{\lambda'}^{-1}(v)-g_{\lambda'}^{-1}(u)| < W,$
$| D(u,v) | > \eta,$ and
$\phi(u,v) > \eta$  for all $\lambda' \in [0,1]$ satisfying $|\lambda'-\lambda| < \eta.$

For all $\delta \in [0,2W]$, define
\[
\theta(\delta) = \min_{x\in [-W+\delta,W]}(\Omega(x)-\Omega(x-\delta)).
\]
Note that $\theta(\delta)>0$ for $\delta>0$ and that $\theta$ is non-decreasing.

Let us find $a<b$ such that $|a-\lambda|<\eta$ and $|b-\lambda|<\eta.$
Then, for any $(u,v)\in A,$ we have
\begin{align*}
\int_a^b \mathrm{d}\lambda D&(u,v)^2 \smthker (f_0^{-1}(v)-g_0^{-1}(u)+\lambda D(u,v))
\\ &=
 D(u,v) \big[\Omega (f_0^{-1}(v)-g_0^{-1}(u)+b D(u,v)) - \Omega (f_0^{-1}(v)-g_0^{-1}(u)+ a D(u,v))\big]
\\ & \ge |D(u,v)|\, \theta((b-a) |D(u,v)|)
\\ & \ge \eta \, \theta((b-a) \eta) .
\end{align*}
By the Fubini theorem, this contradicts our above established claim~\eqref{eqn:claimedformula}.
\end{IEEEproof}

Let us define $D_W = \{ (u,v): |f^{-1}(v)-g^{-1}(u)| < W \}$
and let $B_\epsilon (u)$ denote the open interval centered at $u$ of length $2\epsilon.$

\begin{lemma}\label{lem:SPGCfits}
Let $f,g$ be a CFP and assume the SPGC and the interval support condition.
For all $v\in (0,1)$, there exists $u\in (0,1)$ and $\epsilon > 0$
such that $B_\epsilon(u) \times B_\epsilon(v)  \subset D_W.$
\end{lemma}
\begin{IEEEproof}
Let $v\in (0,1)$ and define $x_m = \frac{1}{2}(f^{-1}(v+)+f^{-1}(v-)).$
We must have $f^{-1}(v+)-f^{-1}(v-) < 2W$ since, otherwise, we obtain
$\altPhiSI(\smthker; f,g;x_m,x_m) = 0$
which, by Lemma~\ref{lem:spatialintegration}, contradicts the SPGC.

To be more precise, 
for $z\in \reals$, let us define
\begin{align*}
T_1(z) &= C_1(z,z) \cap \{ |x-y|<W\} = \{ (x,y): x>z,y\le z, x-y <W \},
\\
T_2(z) &=C_2(z,z)  \cap \{ |x-y|<W\}  =\{ (x,y): x\le z,y> z, y-x <W \}.
\end{align*}
By~\eqref{eqn:altPhidef} (see also~\eqref{eqn:altphiform}), the SPGC implies that the $\mathrm{d}f \mathrm{d}g$ measure
of  at least one of $T_1(x_m)$ and $T_2(x_m)$ is strictly positive.
We shall assume that the measure of  $T_1(x_m)$ is positive,
and the other case can be handled similarly.
It follows from monotonicity of $f$ and $g$ that there exists $u \in (0,1)$ and $\epsilon$ sufficiently 
small such that $g^{-1}(B_\epsilon(u)) \times f^{-1}((v,v+\epsilon)) \subset T_1(x_m).$
We then have $g^{-1}(B_\epsilon(u)) \subset (x_m-W,x_m]$ and, for $\epsilon$ small enough, 
$f^{-1}((v-\epsilon,v))\subset (x_m-W,x_m]$, which gives
$B_\epsilon(u) \times B_\epsilon(v)  \subset D_W.$
\end{IEEEproof}

By Lemma~\ref{lem:displaceflat}, we  have that, if $f_0,g_0$ and $f_1,g_1$ are CFPs
and the SPGC and interval support condition holds, then
\begin{align}\label{eqn:SPGCsetcond}
\mu
\{
(u,v) : | f_\lambda^{-1}(v)-g_\lambda^{-1}(u) | < W,  D(u,v) \neq 0
\}
=0.
\end{align}
We claim that this implies that $f_1^{-1}(v) - f_0^{-1}(v)$ is essentially constant.
Similarly, we have $g_1^{-1}(u) - g_0^{-1}(u)$ is essentially constant.
Moreover, these two constants are equal.

\begin{lemma}\label{lem:essconst}
Assume the SPGC  and the interval support condition and that $f_1,g_1$ and $f_0,g_0$ are CFPs.
Then, $f_1^{-1} - f_0^{-1}$  is essentially constant on $[0,1].$
\end{lemma}
\begin{IEEEproof}
Let us assume that $D_f=f_1^{-1} - f_0^{-1}$ is not essentially constant,
i.e., there exists a real value $s$ so that
$| \{   v: D_f(v) >s \}| \in (0,1)$
and
$| \{   v: D_f(v) \le s \}| \in (0,1)$.
Then, there  exists a value $v^* \in (0,1)$ that is in the support of both sets,
i.e.,
 for any $\epsilon >0$ we have 
$| \{   v: D_f(v) >s \}\cap B_\epsilon (v^*)| > 0$ and
 $| \{   v: D_f(v) \le s \}\cap B_\epsilon (v^*)| >0.$ 
 
By Lemma~\ref{lem:SPGCfits},
there exists a $u\in (0,1)$ and $\epsilon > 0$
such that $B_\epsilon(u) \times B_\epsilon(v^*)  \subset D_W.$
By definition of $v^*$, there is a positive constant $\eta > 0$ such that
$\int_{B_\epsilon(v^*) } {\mathrm d}v \, |D(u,v)| > \eta$ for all $u\in B_\epsilon(u)$,
which now contradicts~\eqref{eqn:SPGCsetcond}.
This completes the proof.
\end{IEEEproof}

\begin{proposition}\label{prop:unicity}
Assume the SPGC and the interval support condition and that $f_1,g_1$ and $f_0,g_0$ are interpolating monotonic CFPs.
Then, there exists $m$ such that, for almost all $x$, we have
$f_1(x)=f_0(x+m)$ and $g_1(x)=g_0(x+m).$
\end{proposition}

\begin{IEEEproof}
By Lemma~\ref{lem:essconst}, 
there exists $m$ such that $f_1^{-1}(v) - f_0^{-1}(v) =m$ 
for almost all  $v\in [0,1].$
Similarly, there exists
$m'$ such that $g_1^{-1}(u) - g_0^{-1}(u) =m'$ 
for almost all  $u\in [0,1].$
It follows that $D(u,v) = m-m'$ for almost all $(u,v) \in [0,1]^2.$
It now follows from Lemma~\ref{lem:SPGCfits} and~\eqref{eqn:SPGCsetcond}
that $m=m'.$
\end{IEEEproof}

Even though we have stated and proved the results for CFPs, under the assumptions
of this section CFPs are actually FPs.
\begin{lemma}
If $f,g$ is a CFP and $h_f,h_g$ satisfies the strictly positive gap condition
and the interval support condition holds, then $f,g$ is a FP.
\end{lemma}
\begin{IEEEproof}
If the SPGC and the interval support condition hold then $f^\smthker$ and $g^\smthker$ are strictly increasing
wherever they take values in $(0,1).$
This implies that $f,g$ must be a FP
(see \cite{KRU12} for further detail).
\end{IEEEproof}

\section{Illustrations}\label{SectionApplications}

In this work, we have shown  (Proposition~\ref{thm:existence} and Propositions~\ref{propConv},~\ref{prop:CFPmin}, and~\ref{prop:unicity}) that under some conditions, the potential functional $\mathcal{W}$ is displacement convex and that its minimizer exists and is unique up to translation. These conditions are the strictly positive gap condition, $C_\smthker<\infty$, and the interval support condition. In this section, we apply these results on different scalar systems when these conditions hold. In particular, for the applications we consider, we use the even uniform window with $W=\frac{1}{2}$ which implies the two latter conditions. We illustrate for each application that the strictly positive gap condition holds.

To check the SPGC one can directly look at $\phi(h_f, h_g; u,v)$, but there is also a simpler way to check the condition. Indeed, we already remarked that for fixed $u$ the potential is minimized by setting $v=h_f(u)$. Therefore, 
\begin{align*}
\phi(h_f, h_g; u, v) \geq \phi(h_f, h_g; u, h_f(u)) = A(h_f, h_g; u).
\end{align*}
So the SPGC is valid as long as the signed area $A(h_f, h_g; u) > 0$ for $u\in (0, 1)$. Similarly, for fixed $v$, the potential is minimized by setting $u=h_g(v)$. Thus, 
\begin{align*}
\phi(h_f, h_g; u, v) \geq \phi(h_f, h_g; h_g(v), v) = \tilde{A}(h_f, h_g; v),
\end{align*}
where 
\begin{align*}
\tilde{A}(h_f, h_g; v) = \int_0^v dv^\prime (h_f^{-1}(v^\prime) - h_g(v^\prime))
\end{align*}
is the alternative signed area bounded between the two EXIT curves and the horizontal axis at the origin and at height $v$. The SPGC is valid as long as 
$\tilde{A}(h_f, h_g; v) > 0$ for $v\in (0,1)$.

Clearly, when $\phi(h_f, h_g; 1, 1)=0$ as assumed in this paper, we also have 
$A(h_f, h_g; 1) = \tilde{A}(h_f, h_g; 1)$. 

\subsection{LDPC Code Ensembles on the BEC}

\begin{figure}[tb!]
\centering
\includegraphics[draft=false,scale=0.27]{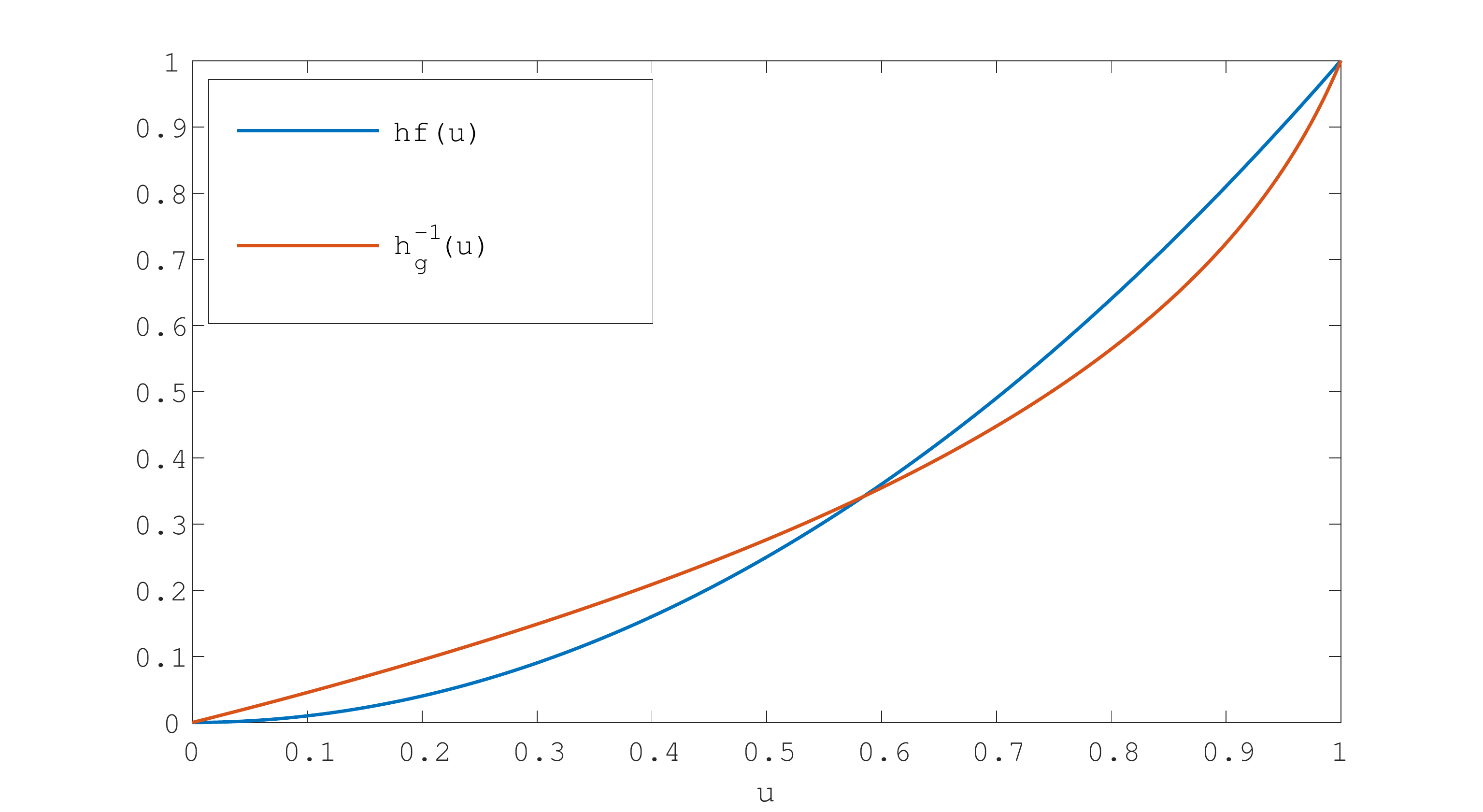}
\caption{We plot the EXIT curves $h_f(u)$ and $h_g^{-1}(u)$ for $u\in[0,1]$ for the $(3,6)$-regular LDPC ensemble with transmission over the BEC(0.4881). We note that the signed area between the curves is equal to zero.}
\label{fig:exitBEC}
\end{figure}

\begin{figure}[tb!]
\centering
\includegraphics[draft=false,scale=0.27]{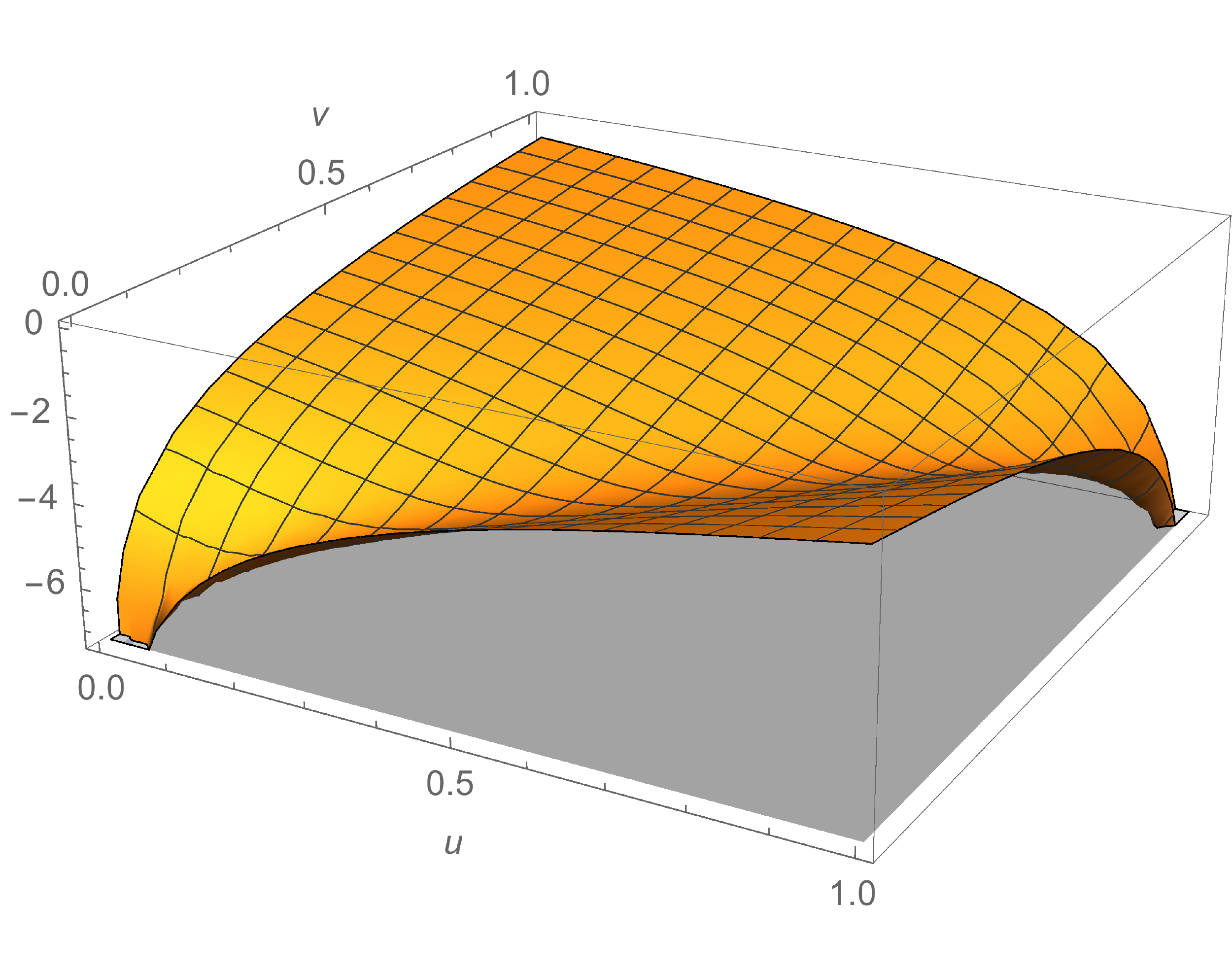}
\caption{We consider the $(3,6)$-regular LDPC ensemble with transmission over the BEC($\epsilon$). We plot $\phi(h_f,h_g;u,v)$ for $(u,v)\in[0,1]^2$ in log scale when $\epsilon=\epsilon_{\text{\tiny MAP}}=0.4881$. We can see that $\phi(h_f,h_g;u,v)>0$ for $(u,v)\in(0,1)^2$ and $\phi(h_f,h_g;0,0)=\phi(h_f,h_g;1,1)=0$.}
\label{fig:phiBEC}
\end{figure}

We demonstrate our results on the $(3,6)$-regular spatially coupled LDPC code ensemble when transmission 
takes place over the BEC($\epsilon$). For this ensemble, we have the (unscaled) uncoupled 
DE equations $\mathrm{x} = \epsilon \mathrm{y}^{2}$ and $\mathrm{y} = 1-(1-\mathrm{x})^{5}$. 
We already showed how to perform the right scaling $\mathrm{x} = \mathrm{x}_{\text{\tiny MAP}} v$ and $\mathrm{y} = \mathrm{y}_{\text{\tiny MAP}} u$;
asking that $(u,v)=(1,1)$ is a fixed point and $A(h_f,h_g;1)=0$ we find 
 $\mathrm{y}_{\text{\tiny MAP}}=0.941$, $\mathrm{x}_{\text{\tiny MAP}} = 0.432$ and $\epsilon=\epsilon_{\text{\tiny MAP}}=0.4881$. Replacing these numbers in the expression of the potential function (see Section 
~\ref{subsec:potfunctions}) we find $\phi(u,v)$. Fig.~\ref{fig:exitBEC} and~\ref{fig:phiBEC} illustrate the corresponding EXIT curves and the potential that is seen to satisfy the SPGC.

\subsection{Generalized LDPC Codes}

We consider a generalized LDPC (GLDPC) code, where the check node constraints are given by a primitive BCH code with minimum distance $d=2e+1$ 
(see \cite{jian2012approaching} for more information). 
We consider the code with degree-2 variable nodes and degree-$n$ check nodes, with transmission over the BEC($\epsilon$). 
The (unscaled) uncoupled DE equations are \cite{yedla2014simple} 
\begin{align*}
\begin{cases}
\mathrm{x} =\epsilon \mathrm{y},\\
\mathrm{y} = \sum_{i=e}^{n-1} {n-1 \choose i} \mathrm{x}^i(1-\mathrm{x})^{n-i-1}.
\end{cases}
\end{align*}
Set $\epsilon=\epsilon_{\text{\tiny MAP}}$ and $\mathrm{y} = \mathrm{y}_{\text{\tiny MAP}} u$, $\mathrm{x} = \mathrm{x}_{\text{\tiny MAP}} v$. We then get the scaled equations~\eqref{simple-DE}, namely $v= h_f(u)$, $u=h_g(v)$ 
\begin{align*}
\begin{cases}
h_f(u) = \epsilon_{\text{\tiny MAP}} \mathrm{x}_{\text{\tiny MAP}}^{-1} \mathrm{y}_{\text{\tiny MAP}} u,\\
h_g(v) = \mathrm{y}_{\text{\tiny MAP}}^{-1} 
\sum_{i=e}^{n-1} {n-1 \choose i} \mathrm{x}_{\text{\tiny MAP}}^i\mathrm{v}^i(1-\mathrm{x}_{\text{\tiny MAP}} v)^{n-i-1}.
\end{cases}
\end{align*}
The normalization condition $h_f(1) = h_g(1) =1$ and the condition 
$\tilde A(h_f, h_g; 1) =0$ completely determine $\epsilon_{\text{\tiny MAP}}$, 
$\mathrm{x}_{\text{\tiny MAP}}$, and $\mathrm{y}_{\text{\tiny MAP}}$. The potential function and (alternative) signed area are given by 
\begin{align*}
\phi(u,v) &= \frac{\mathrm{x}_{\text{\tiny MAP}}u^2}{2\epsilon_{\text{\tiny MAP}} \mathrm{y}_{\text{\tiny MAP}}} + \int_0^v\mathrm{d}v' \, h_g^{-1}(v') -uv.\\
\tilde{A}(h_f,h_g;v) &= \frac{\mathrm{x}_{\text{\tiny MAP}}v^2}{2\epsilon_{\text{\tiny MAP}} \mathrm{y}_{\text{\tiny MAP}}}
-\frac{1}{\mathrm{y}_{\text{\tiny MAP}}} \sum\limits_{i=e}^{n-1} \sum\limits_{m=0}^{n-i-1} {n-1 \choose i} {n-i-1 \choose m} \frac{\mathrm{x}_{\text{\tiny MAP}}^{m+i} v^{m+i+1}}{m+i+1}.
\end{align*}

\begin{figure}
\centering
\includegraphics[draft=false,scale=0.27]{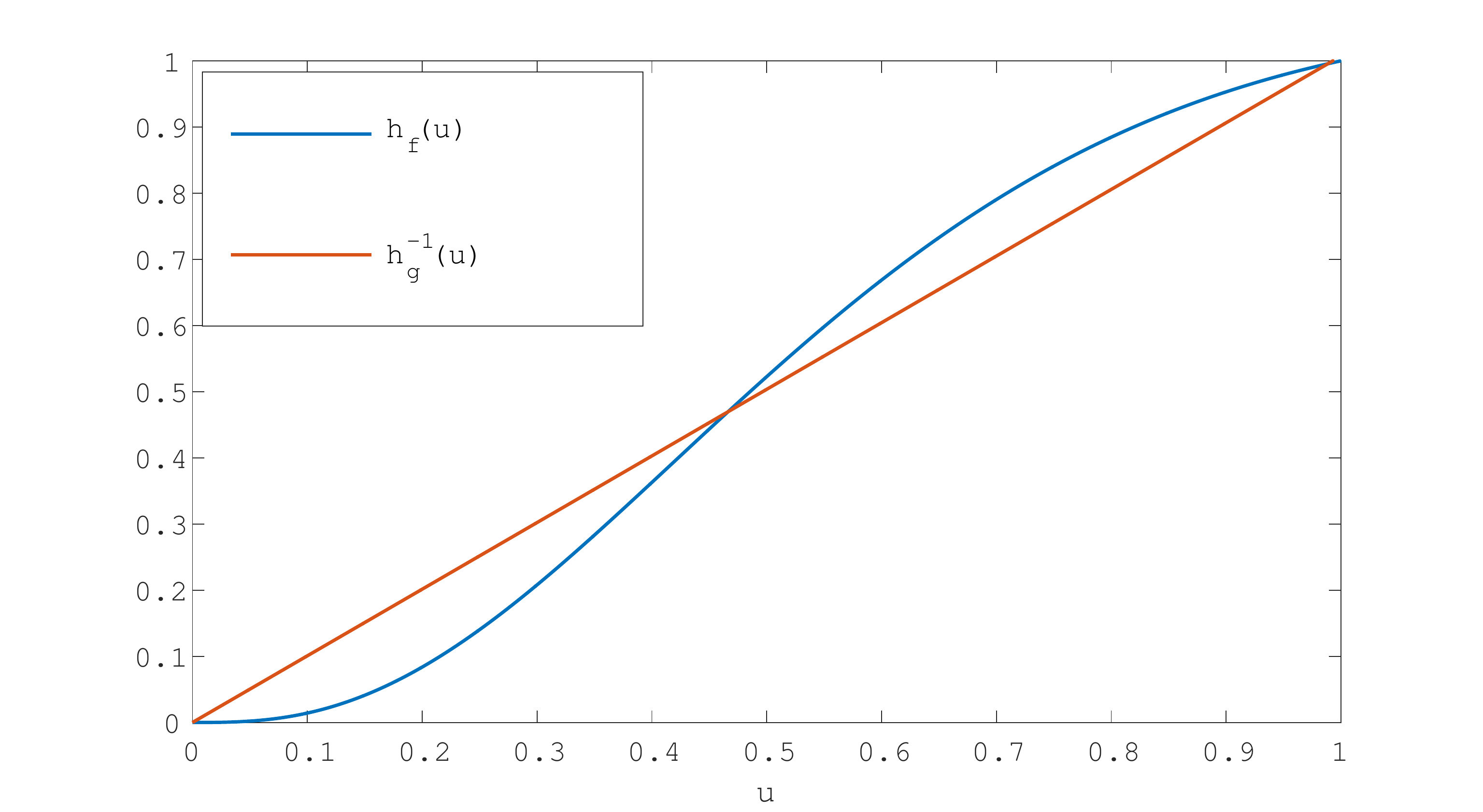}
\caption{We plot the EXIT curves $h_f(u)$ and $h_g^{-1}(u)$ for $u\in[0,1]$ for the GLDPC code with $n=15$ and $e=3$, when transmission takes place over the BEC(0.3901). We note that the signed area between the curves is equal to zero.}
\label{fig:exitGLDPC}
\end{figure}

\begin{figure}
\centering
\includegraphics[draft=false,scale=0.27]{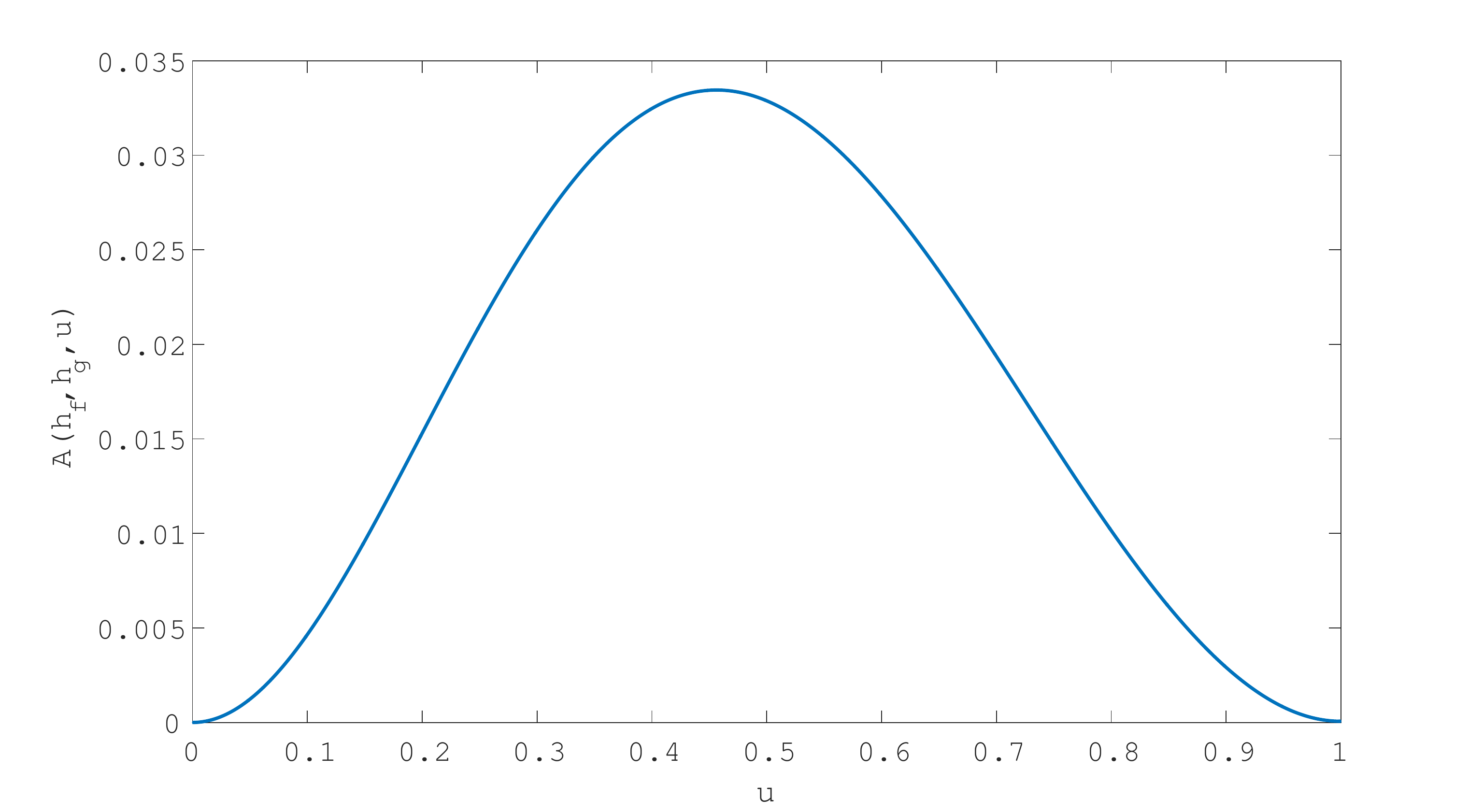}
\caption{We consider the GLDPC code with $n=15$ and $e=3$, with transmission over the BEC($\epsilon$). We plot $\tilde{A}(h_f,h_g;v)$ for $v\in [0,1]$ when the channel parameter is $\epsilon=\epsilon_{\text{\tiny MAP}}=0.3901$.}
\label{fig:AGLDPC}
\end{figure}

The EXIT curves and signed area are illustrated in Fig.~\ref{fig:exitGLDPC}. 
and Fig.~\ref{fig:AGLDPC} for the GLDPC code with $n=15$ and $e=3$. This corresponds to 
$\mathrm{x}_{\text{\tiny MAP}} = 0.3670$, $y_{\text{\tiny MAP}} = 0.9342$,
$\epsilon_{\text{\tiny MAP}}=0.3901$. 
Clearly, the SPGC condition is satisfied.

\subsection{The Gaussian Approximation}

There are various forms of the Gaussian approximation \cite{chung2001analysis}, \cite{chung2001capacity}, \cite{chung2000construction} used to simplify the analysis of coding systems with transmission over binary memoryless symmetric (BMS) channels. Here, we consider a variant developed in \cite{chung2001capacity}, \cite{chung2000construction}.  
This method approximates the densities of the log-likelihood ratio (LLR) messages exchanged in the decoding graph with {\it symmetric} Gaussian densities; that is, densities of the form $\mathtt{x}(\alpha) = 1/\sqrt{2\pi\sigma^2}\exp(-\frac{(\alpha-m)^2}{2\sigma^2})$ with the property $\sigma^2 = 2m$. We also approximate the BMS channel $\mathtt{c}$ with a binary-input Gaussian additive white noise (BIGAWN) channel with parameter $\sigma^2$ and with the same entropy $H(\mathtt{c})$ as the original channel $\mathtt{c}$. This makes the analysis one-dimensional and has been shown to serve as a good approximation.

The Gaussian approximation allows us to track the evolution of decoding by tracking the entropies of the LLR messages. 
Let $\psi(m)$ denote the entropy of a symmetric Gaussian density of mean $m$ \cite{richardson2008modern}. In particular, it can be expressed as 
\begin{align*}
\psi(m)=\frac{1}{\sqrt{4\pi m}}\int_\reals \mathrm{d}z \, e^{-\frac{(z-m)^2}{4m}}\log_2(1+e^{-z}).
\end{align*} 
Note that $\psi(+\infty)=0$ and $\psi(0)=1$.
We consider the $(3,6)$-regular LDPC code ensemble with transmission over the BMS. 

The (unscaled) uncoupled DE equations are
\begin{align*}
\begin{cases}
\mathrm{x} = \psi( \psi^{-1}( H(\mathtt{c})) + 2 \psi^{-1}( \mathrm{y} ) ),\\
\mathrm{y} = 1 - \psi( 5\psi^{-1} (1-\mathrm{x}) ).
\end{cases}
\end{align*}
We define $m_{\text{\tiny MAP}}$ as the value of $\psi^{-1}( H(\mathtt{c}))$ at the MAP threshold and set $\psi^{-1}( H(\mathtt{c}))=m_{\text{\tiny MAP}}$ and $\mathrm{x} = \mathrm{x}_{\text{\tiny MAP}} v$, $\mathrm{y} = \mathrm{y}_{\text{\tiny MAP}} u$. We then get the scaled equations~\eqref{simple-DE}, namely $v= h_f(u)$, $u=h_g(v)$ 
\begin{align*}
\begin{cases}
h_f(u) = \mathrm{x}_{\text{\tiny MAP}}^{-1} \psi( m_{\text{\tiny MAP}} + 2 \psi^{-1}( \mathrm{y}_{\text{\tiny MAP}} u ) ),\\
h_g(v) = \mathrm{y}_{\text{\tiny MAP}}^{-1} - \mathrm{y}_{\text{\tiny MAP}}^{-1} \psi( 5\psi^{-1} (1-\mathrm{x}_{\text{\tiny MAP}} v) ).
\end{cases}
\end{align*}
The normalization condition $h_f(1) = h_g(1) =1$ and the condition 
$\tilde A(h_f, h_g; 1) =0$ completely determine $m_{\text{\tiny MAP}}$, 
$\mathrm{x}_{\text{\tiny MAP}}$, and $\mathrm{y}_{\text{\tiny MAP}}$. The potential function is given by 
\begin{align*}
\phi(u,v) =& u(\mathrm{x}_{\text{\tiny MAP}}^{-1}-v)
+ \mathrm{x}_{\text{\tiny MAP}}^{-1}\int_0^u\mathrm{d}u' \, \psi\Big(\frac{1}{5}\psi^{-1}(1-\mathrm{y}_{\text{\tiny MAP}}u') \Big)
+ \mathrm{y}_{\text{\tiny MAP}}^{-1}\int_0^u\mathrm{d}v' \, \psi\Big(\frac{1}{2}\psi^{-1}(\mathrm{x}_{\text{\tiny MAP}}v')-\frac{1}{2}m_{\text{\tiny MAP}} \Big).
\end{align*}
%\begin{align*}
%\tilde{A}(h_f,h_g;v) & = -\mathrm{y}_{\text{\tiny MAP}}^{-1}v\\
%&+\mathrm{y}_{\text{\tiny MAP}}^{-1}\int_0^v\mathrm{d}v' \, \Big\{\psi\Big(\frac{1}{2}\psi^{-1}(\mathrm{x}_{\text{\tiny MAP}}v')-\frac{1}{2}m_{\text{\tiny MAP}} \Big) 
%\\ &
%+ \psi\Big(5\psi^{-1}(1-\mathrm{x}_{\text{\tiny MAP}}v') \Big) \Big\}.
%\end{align*}
A plot of the EXIT curves and potential function yields curves that are very similar to the case of the BEC (see e.g. Figs~\ref{fig:exitBEC},~\ref{fig:phiBEC}).

\subsection{Compressive Sensing}

Consider a signal vector $\mathbf{s}$ of length $n$ where the components are i.i.d. copies of a random variable $S$. We assume that $\mathbb{E}[S^2]=1$ and that each component of $\mathbf{s}$ is corrupted with Gaussian noise $\mathcal{N}(0,\sigma^2=1/\mathtt{snr})$. We take $m$ measurements of the signal and assume that the measurement matrix has i.i.d. Gaussian components $\mathcal{N}(0,1/\sqrt{n})$. The measurement ratio is defined by $\delta=m/n$. Here we are interested in state evolution \cite{donoho2012information}, which tracks the mean square error of the approximate message-passing (AMP) estimator (for the signal) . Given an $\mathtt{snr}$ that is large enough, the parameter $\delta$ is kept fixed as $n$ gets large. 

The state evolution fixed point equations read
\begin{align}\label{state-ev}
\begin{cases}
\mathrm{y} = (\frac{1}{\mathtt{snr}}+\frac{\mathrm{x}}{\delta})^{-1},\\
\mathrm{x} = \mathtt{mmse}(\mathrm{y}),
\end{cases}
\end{align}
where the minimum mean square error function $\mathtt{mmse}$ is defined as follows. Let $Y = \sqrt{\mathtt{snr}}S+Z$ where $Y$ is a scalar output and $Z\sim \mathcal{N}(0,1)$ and let $\hat{S}(Y,\mathtt{snr}) = \mathbb{E}_{S|Y}[S|Y]$. Then 
$\mathtt{mmse}(\mathtt{snr}) = \mathbb{E}_{S,Y}[(S-\hat{S}(Y,\mathtt{snr}))^2]$.
In the equations above, when we initialize with $x^{(0)}=1$,  $x^{(t)}$ is the average mean square error of the AMP estimator at iteration $t$.

We now put this system of equations in the form~\eqref{simple-DE}. Here, there is no trivial fixed point $\mathrm{x}=\mathrm{y}=0$; however, the picture is very similar to LDPC coding-like systems considered above. The role of the ``trivial" fixed point is played by a fixed point $\mathrm{x}_*, \mathrm{y}_*$ obtained by initializing state evolution with $\mathrm{x}=0$. Given the $\mathtt{snr}$, for $\delta$ below the algorithmic threshold, this is the only fixed point, and for $\delta$ above this threshold, one finds three solutions (besides $\mathrm{x}_*, \mathrm{y}_*$ which is stable, there are an unstable and a stable fixed point). Set $\mathrm{x}^\prime = \mathrm{x}-\mathrm{x}_*$ and $\mathrm{y}^\prime= \mathrm{y}- \mathrm{y}_*$. Equations~\eqref{state-ev} become
\begin{align}\label{state-ev-fixed}
\begin{cases}
\mathrm{y}^\prime = - \mathrm{y}_*+(\frac{1}{\mathtt{snr}}+\frac{\mathrm{x}_*+\mathrm{x}^\prime}{\delta})^{-1},\\
\mathrm{x}^\prime = - \mathrm{x}_* +\mathtt{mmse}(\mathrm{y}_* + \mathrm{y}^\prime).
\end{cases}
\end{align}
Note that $\mathrm{x}^\prime= \mathrm{y}^\prime=0$ is a fixed point. We now scale 
$\mathrm{x}^\prime = \mathrm{x}_{\text{\tiny MAP}} v$, $\mathrm{y}^\prime = \mathrm{y}_{\text{\tiny MAP}} u$ where $\mathrm{x}_{\text{\tiny MAP}}$ and $\mathrm{y}_{\text{\tiny MAP}}$ are chosen later on. Then~\eqref{state-ev-fixed} takes the form 
\eqref{simple-DE} with the EXIT curves defined as 
\begin{align}\label{exit-comp-sensing}
\begin{cases}
h_g(v) = - \mathrm{y}_*\mathrm{y}_{\text{\tiny MAP}}^{-1}+(\frac{1}{\mathtt{snr}}+\frac{\mathrm{x}_*+\mathrm{x}_{\text{\tiny MAP}}v}{\delta})^{-1}\mathrm{y}_{\text{\tiny MAP}}^{-1},\\
h_f(u) = - \mathrm{x}_* \mathrm{x}_{\text{\tiny MAP}}^{-1} +\mathtt{mmse}(\mathrm{y}_* + \mathrm{y}_{\text{\tiny MAP}} u)\mathrm{x}_{\text{\tiny MAP}}^{-1}.
\end{cases}
\end{align}
From these, one can compute the potential and the signed areas. Here, we illustrate the signed area. We have 
\begin{align*}
h_g^{-1}(u) = -\mathrm{x}_*\mathrm{x}_{\text{\tiny MAP}}^{-1} + \delta ((\mathrm{y}_* + \mathrm{y}_{\text{\tiny MAP}} u)^{-1} + \mathtt{snr}^{-1}),
\end{align*}
from which it follows that
\begin{align*}
A(h_f, h_g; u) = & \frac{\delta}{\mathrm{x}_{\text{\tiny MAP}}\mathrm{y}_{\text{\tiny MAP}}}\ln(1 +  \frac{\mathrm{y}_{\text{\tiny MAP}}}{\mathrm{y}_*} u)
+
\frac{u \delta }{\mathrm{x}_{\text{\tiny MAP}}\mathtt{snr}}
- \frac{1}{\mathrm{x}_{\text{\tiny MAP}}}
\int_0^u \mathrm{d}u^\prime\, \mathtt{mmse}(\mathrm{y}_* + \mathrm{y}_{\text{\tiny MAP}} u^\prime)
\end{align*}
Finally, we set the signal-to-noise ratio to the value $\mathtt{snr}_{\text{\tiny MAP}}$ defined such that $h_f(1)=h_g(1)=1$ and $A(h_f, h_g; 1)\vert_{\mathtt{snr}_{\text{\tiny MAP}}} = 0$. These conditions also determine 
$\mathrm{x}_{\text{\tiny MAP}}$ and $\mathrm{y}_{\text{\tiny MAP}}$ (note also that these values are a ``non-trivial" stable fixed point). 
A plot of $A(h_f, h_g; u)\vert_{\mathtt{snr}_{\text{\tiny MAP}}}$
at $\mathtt{snr}_{\text{\tiny MAP}}$ yields a curve similar to Fig.~\ref{fig:AGLDPC} that satisfies the SPGC.

\section{Conclusion}

%The performance of a system governed by coupled scalar recursions can be expressed in terms of a potential functional. In this paper we show that, under mild conditions on the system  most of which are necessary for the existence of a FP, the potential is  displacement convex. 
%Under the conditions used in \cite{KRU12} to show existence of spatial FPs we
%show that the displacement convexity is \emph{strict}. This ensures that the pair of profiles minimizing the potential is unique up to translation.

There are some questions that remain open. We have seen in Section~\ref{sectionRearr} that we restrict our search of 
minimizing profiles to the space of increasing profiles. It is not clear in our settings when 
the inequality \eqref{eqnIncRearrB} is strict and so we cannot exclude the existence of a minimizing pair outside the spaces of increasing profiles. 
Another more fundamental open problem comes back to our formulation of the potential.  In applications, it is inherently discrete whereas in our 
analysis, it is convenient to consider the continuum limit approximation of the potential. It would be interesting to see whether this analysis 
can be adapted to the discrete formulation.

\section*{Acknowledgment} 
We thank Vahid Aref and Marc Vuffray for interesting discussions at the early stages of this work.

\appendix
 \label{app}

%The appendix contains the proof of the most general results that depend on taking limits.
The appendix contains proofs of the various limit results that allow the generalization of arguments from the saturated case to the non-saturated case, as well as 
some elementary technical results.
\subsection{Integrability}\label{app:integrability}
\begin{lemma}\label{lem:limintegral}
Let $p$ be an interpolating profile and assume that $C_\smthker < \infty.$ Then,
\[
\lim_{A,B\rightarrow\infty} \int_{-A}^B {\mathrm d}x \, (p^\smthker(x) - p(x))= 0\,.
\]
\end{lemma}
\begin{IEEEproof}
Assume that $C_\smthker < \infty$.
By the evenness of $\smthker$ we have
\begin{align*}
p^\smthker(x) - p(x) & = \int_0^{+\infty}\mathrm{d}y\, w(y)(p(x-y) - p(x))
 + \int_{-\infty}^0\mathrm{d}y\, w(y)(p(x-y) - p(x))
\\ &
= \int_0^\infty {\mathrm d}y\smthker(y)\, (p(x-y)+p(x+y)-2p(x))\,.
\end{align*}
Applying the Fubini theorem, we have
\[
 \int_{-A}^B {\mathrm d}x \, (p^\smthker(x) - p(x)) = 
 \int_0^\infty {\mathrm d}y\smthker(y)\, (D(B,y)-D(-A,y)),
\]
where we introduce the notation
\begin{align*}
D(B,y) &=
\int_{B}^{B+y} {\mathrm d}z\, p(z) 
-\int_{B-y}^B {\mathrm d}z\, p(z)  
\\& =
\int_{B-y}^B {\mathrm d}z\,(1- p(z)) 
-\int_{B}^{B+y} {\mathrm d}z\,(1- p(z))\,. 
\end{align*}
%Then, by taking $p = H_f \circ f$ we complete the proof.
From these two expressions we obtain the two bounds
\begin{align*}
|D(B,y)| &\le
y \sup_{z < B+y}  p(z),
\\
|D(B,y)|& \le
y \sup_{z > B-y} (1- p(z))\,. 
\end{align*}
Letting $K>0$ be arbitrary, we have
\begin{align*}
\bigl|\int_0^{\infty}{\mathrm d}y \,\smthker(y) D(B,y)\bigr|
&\le
\int_0^{\infty}{\mathrm d}y \,\smthker(y) |D(B,y)|
\\& 
\le
\int_K^{\infty}{\mathrm d}y \,\smthker(y) y + \sup_{z > B-K} (1- p(z)) \int_0^{K}{\mathrm d}y \,\smthker(y) y  
\end{align*}
Since $C_\smthker< \infty$ we see, by choosing $K=B/2$, that we have
\begin{align*}
\lim_{B\rightarrow\infty}
\int_0^{\infty}{\mathrm d}y \,\smthker(y) D(B,y)
=0\,.
\end{align*}

Similarly, we have
\begin{align*}
&\bigl|\int_0^{\infty}{\mathrm d}y \,\smthker(y) D(-A,y)\bigr|
\le
\int_K^{\infty}{\mathrm d}y \,\smthker(y) y + \sup_{z <-A+K} p(z) \int_0^{K}{\mathrm d}y \,\smthker(y) y  ,
\end{align*}
which, by choosing $K=A/2$, gives
\begin{align*}
\lim_{A\rightarrow\infty}
\int_0^{\infty}{\mathrm d}y \,\smthker(y) D(-A,y)
=0\,.
\end{align*}
\end{IEEEproof}

\subsection{Basic Bounds}\label{app:basicbounds}

We begin with some approximation limits.

\begin{lemma}\label{lem:klimrearr}
Let $p$ be an interpolating profile (i.e., one satisfying~\eqref{eqn:RearrLimits})
and assume that $C_\smthker < \infty.$ Then
\begin{align}
\lim_{K\rightarrow \infty}
\int_K^\infty \mathrm{d}x\,
(1-\cut{p}{K}^\smthker(x)) &=0\label{eqn:tailconv0}
\\
\lim_{K\rightarrow \infty}
\int^{-K}_{-\infty} \mathrm{d}x\,
\cut{p}{K}^\smthker(x) &=0\label{eqn:tailconv0left}
\\
\lim_{K\rightarrow \infty}
\int_{-K}^{K} \mathrm{d}x\,
|p^\smthker(x)-\cut{p}{K}^\smthker(x)| &=0\label{eqn:smthconv0}
\end{align}
\end{lemma}
\begin{IEEEproof}
Define 
\[
\xi(K) = \sup_{x\ge K}\{ 1-p(x),p(-x) \},
\]
and note that $\lim_{K\rightarrow \infty} \xi(K)=0.$  We have
\begin{align*}
1-\cut{p}{K}(x)&\le \indicator{x\le K/2}+\xi(K/2)\indicator{K/2<x\le K}\\
 &
\le \indicator{x\le K/2}+\xi(K/2)\indicator{x\le K},
\end{align*}%&=(1-\xi(K/2))\indicator{x\le K/2}+\xi(K/2)\indicator{x\le K}
from which we obtain (using changes of variables)
\begin{align*}
\int_K^\infty \mathrm{d}x\,
(1-\cut{p}{K}^\smthker(x)) 
&=
\int_K^\infty \mathrm{d}x\, 
\int_\reals \mathrm{d}y\, \smthker(x-y) (1-\cut{p}{K}(y)) 
\\
&\le 
\int_K^\infty \mathrm{d}x\, 
\big(
\Omega (K/2-x)
+
\Omega (K-x) \xi(K/2)
 \big),
\\
&=
 V(-K/2)+V(0)\xi(K/2)
\end{align*}
and~\eqref{eqn:tailconv0} now follows.
The inequality~\eqref{eqn:tailconv0left} can be shown similarly by first noting that 
\begin{align*}
\cut{p}{K}(-x)&%\le (1-\xi(K/2))\indicator{x\le K/2}+\xi(K/2)\indicator{x\le K}
\le \indicator{x\le K/2}+\xi(K/2)\indicator{x\le K},
\end{align*}
and writing
\begin{align*}
\int_{-\infty}^{-K}\mathrm{d}x\,\cut{p}{K}^\smthker(x)&=\int_K^{\infty}\mathrm{d}x\,\cut{p}{K}^\smthker(-x)
=\int_K^{\infty}\mathrm{d}x\int_{-\infty}^{K/2}\mathrm{d}y\,w(y-x)\cut{p}{K}(-y).
\end{align*}
Using again changes of variables and the upper bound on $\cut{p}{K}(-y)$, we find that $\int_{-\infty}^{-K}\mathrm{d}x\,\cut{p}{K}^\smthker(x) \leq V(-K/2)+V(0)\xi(K/2)$, which proves~\eqref{eqn:tailconv0left}.

Now we show~\eqref{eqn:smthconv0}.
We have
\begin{align*}
|p^\smthker(x)-\cut{p}{K}^\smthker(x)| 
& \le
\int_{-\infty}^{-K} \mathrm{d}y\,
\smthker(x-y) p(y)
+
\int_K^{\infty} \mathrm{d}y\,
\smthker(x-y) (1-p(y))
\\& \le
\xi(K)\bigl(
\Omega(-x-K)
+
\Omega(x-K)
\bigr)
\end{align*}
%ve used the definition of $\xi(K)=\sup_{x\ge K}\{ 1-p(x),p(-x) \}$. 
from which we obtain
\begin{align*}
\int_{-K}^{K} \mathrm{d}x\,
|p^\smthker(x)-\cut{p}{K}^\smthker(x)| 
\le 2 V(0) \xi(K),
\end{align*}
and
\eqref{eqn:smthconv0} follows.
\end{IEEEproof}

We can now prove Lemma~\ref{lem:klimW}. For convenience we restate the lemma.

{\it  (Lemma~\ref{lem:klimW}):}
Let $f$ and $g$ be interpolating profiles 
and assume the PGC and $C_\smthker < \infty$. 
%and ${\cal W}(f,g) < \infty.$
Then
\[
\lim_{K\rightarrow \infty}
\mathcal{W}(\cut{f}{K},\cut{g}{K}) =
\mathcal{W}(f,g)\,.
\] 
\begin{IEEEproof}[Proof of Lemma~\ref{lem:klimW}]
If ${\cal W}(f,g) = \infty$ then the result follows from Lemma~\ref{lem:posfatou} Equ.~\eqref{eqn:Wfatou}.
We assume now that ${\cal W}(f,g) < \infty$ and note that it is then sufficient to show that
\[
\lim_{K\rightarrow\infty}\Biggl(
\mathcal{W}(\cut{f}{K},\cut{g}{K}) -
\int_{-K}^K \mathrm{d}x
\, I_{f,g,\smthker}(x)
\Biggr) =0\,.
\]
The expression between parentheses can be written as
\begin{align*}
  \int_{-K}^K \mathrm{d}x\,
(f^\smthker(x) - \cut{f}{K}^\smthker(x))g(x)
&
+ \int_K^{+\infty}\mathrm{d}x\,\big(\int_0^1du\,h_g^{-1}(u) + \int_{0}^1dv\, h_f^{-1}(v) - \cut{f}{K}^\smthker(x)\big)
\\ &
=
\int_{-K}^K \mathrm{d}x\,
(f^\smthker(x) - \cut{f}{K}^\smthker(x))g(x)
+
\int_{K}^\infty \mathrm{d}x\,
(1-\cut{f}{K}^\smthker(x)),\,
\end{align*}
where the last term follows from the fact that $A(h_f, h_g;1)=\phi(h_f,h_g;1,1)=0$.
% for 
% interpolating profiles. 
% as shown in Lemma~\ref{lem:PGC}. 
The result now follows from Lemma~\eqref{lem:klimrearr}.
\end{IEEEproof}

\subsection{Rearrangement\label{app:rearrproof}}

Now, we focus on monotonic profiles.
In particular, we prove the following lemma which is used throughout the paper.
% used
% in the proof of Proposition~\ref{prop:mainRearrange}.

\begin{lemma}\label{lem:Cbound}
For any non-decreasing function $h$, we have
\begin{equation}
\int_{\reals} {\mathrm d}x\,|h^w(x)-h(x)|    \le C_\smthker (h(+\infty)-h(-\infty))\,.
\end{equation}
\end{lemma}
\begin{IEEEproof}
First, we note that
\[
h^w(x)-h(x) =
\int_\reals \,\mathrm{d}y (h(x-y)-h(x))\smthker(y),
\]
and we obtain
\begin{align*}
\int_\reals \mathrm{d}x |h^w(x)-h(x)| &\le
\int_\reals \mathrm{d}x \int_\reals \mathrm{d}y
\, |h(x-y)-h(x)|\smthker(y)
\\ &=
\int_\reals \mathrm{d}y \int_\reals \mathrm{d}x
\, |h(x-y)-h(x)|\smthker(y)
\\ &=
\int_\reals \mathrm{d}y\, 
(h(+\infty)-h(-\infty)) |y| \smthker(y)
\\ &=
C_\smthker 
(h(+\infty)-h(-\infty)),
\end{align*} 
where the next-to-last step follows by the layer-cake representation and the monotonicity of $h$. 
\end{IEEEproof}

\subsection{Minimizers}\label{app:CFPs}

In this section, we focus on limit results specific to CFPs.

\begin{lemma}\label{lem:minlimit}
Assume $C_\smthker < \infty$ and let $f,g$ be an interpolating CFP.
Let us define
\begin{align*}
h^K_f &= h_{[\cut{f}{K},\cut{g}{K}^\smthker]}, \quad 
h^K_g = h_{[\cut{g}{K},\cut{f}{K}^\smthker]}. 
\end{align*}
Then
\[
\lim_{K\rightarrow\infty}
{\cal W}(h^K_f,h^K_g;\cut{f}{K},\cut{g}{K})
=
{\cal W}(h_f,h_g;f,g)\,.
\]
\end{lemma}
\begin{IEEEproof}
From~\eqref{eqn:spatialintegration} and~\eqref{eqn:altphiform}, for any CFP, we have
\begin{align*}
\int_\reals & {\mathrm d}x \, \phi(h_f, h_g; g(x), f(x))
= \iint_{\reals^2} \mathrm{d}f(x) \mathrm{d}g(y) |x-y| \Omega(-|x-y|)
\end{align*}
and, since $\lim_{x\rightarrow -\infty} x\Omega(x) =0,$ we clearly have
\begin{align*}
\lim_{K\rightarrow\infty}
\iint_{\reals^2} \mathrm{d}\cut{f}{K}(x) \mathrm{d}\cut{g}{K}(y) |x-y| \Omega(-|x-y|) 
 =
\iint_{\reals^2} \mathrm{d}f(x) \mathrm{d}g(y) |x-y| \Omega(-|x-y|)
\end{align*}
Thus, it only remains to show that
\begin{align*}
\lim_{K\rightarrow\infty}
\int_\reals (\cut{f}{K}(x)-\cut{f}{K}^\smthker(x)) \cut{g}{K}(x) \, {\mathrm d}x 
= 
\int_\reals (f(x)-f^\smthker(x))g(x) \, {\mathrm d}x \,.
\end{align*}
By Lemma~\ref{lem:Cbound}, we have
\begin{align*}
\lim_{K\rightarrow\infty}
\int_{\reals\backslash [-K,K]} (f(x)-f^\smthker(x))g(x) \, {\mathrm d}x \,=0.
\end{align*}
and by Lemma~\ref{lem:klimrearr} Equ.~\eqref{eqn:tailconv0}, we have
\begin{align*}
\lim_{K\rightarrow\infty}
\int_{\reals\backslash [-K,K]}  (\cut{f}{K}(x)-\cut{f}{K}^\smthker(x)) \cut{g}{K}(x)  \, {\mathrm d}x \,=0.
\end{align*}
The result now follows from Lemma~\ref{lem:klimrearr} Equ.~\eqref{eqn:smthconv0}.
\end{IEEEproof}

\subsection{Second derivative}\label{app:second-derivative}
We recall from the proof of Proposition~\ref{propConv} that, for saturated profiles,
\begin{align*}
 \mathcal{W}(f_\lambda, g_\lambda) = L(f_\lambda, g_\lambda) + \int_\reals d\mathrm{x} (1- f^w_\lambda(x)) g_\lambda(x).
\end{align*}
The representation used in Lemma~\ref{lem:convCrossterm} for the second term is equivalent to
\begin{align*}
 \int_\reals \mathrm{d}x\, (1- f^w_\lambda(x)) g_\lambda(x) 
 &= \iint_{[0,1]^2} \mathrm{d}u \mathrm{d}v\, V((1-\lambda)(f_0^{-1}(v) - g_0^{-1}(u)) 
 + \lambda(f_1^{-1}(v) - g_1^{-1}(u)))
 \nonumber \\ &
 = \iint_{[0,1]^2} \mathrm{d}u \mathrm{d}v\, V(f_0^{-1}(v) - g_0^{-1}(u) +\lambda D(u,v))\, .
\end{align*}
Moreover, we saw in Lemma~\ref{lem:Llinear} that $L(f_\lambda, g_\lambda)$ is affine in $\lambda$. So, using $V^{\prime\prime}(x) = w(x)$, we 
immediately get 
\begin{align*}
 \frac{\mathrm{d}^2}{\mathrm{d}\lambda^2}\mathcal{W}(f_\lambda, g_\lambda) 
  &
 =
 \iint_{[0,1]^2} \mathrm{d}u \mathrm{d}v\, D(u,v)^2 w(f_0^{-1}(v) - g_0^{-1}(u) +\lambda D(u,v))
 \nonumber \\ &
 = \iint_{[0,1]^2} \mathrm{d}u \mathrm{d}v\, D(u,v)^2 w(f_\lambda^{-1}(v) - g_\lambda^{-1}(u))\, .
\end{align*}

\bibliographystyle{IEEEtran}
\bibliography{BIBfile}
%,lth,lthpub}

\end{document}